\documentclass[]{iopart}
\usepackage{iopams}
\usepackage{setstack}
\usepackage{graphicx}
\usepackage{amssymb}

\begin{document}

\newcommand{\argmin}[0]{\ensuremath{\mathrm{argmin}}}
\newcommand{\Heven}[0]{\ensuremath{H_{\mathrm{E}}}}
\newcommand{\Hodd}[0]{\ensuremath{H_{\mathrm{O}}}}
\newcommand{\UCN}[0]{\ensuremath{U_{\mathrm{CN2}}}}
\newcommand{\UPADE}[0]{\ensuremath{U_{\mathrm{CN4}}}}
\newcommand{\ket}[1]{\ensuremath{|{#1}\rangle}}
\newcommand{\bra}[1]{\ensuremath{\langle{#1}|}}

\title{Time evolution of Matrix Product States}

\date{\today}

\author{Juan Jos\'e Garc{\'\i}a-Ripoll}
\address{Max-Planck-Institut f\"ur Quantenoptik,
  Hans-Kopfermann-Str. 1,\\
  Garching b. M\"unchen, D-85748, Germany}
\ead{juanjose.garciaripoll@gmail.com}

\begin{abstract}
  In this work we develop several new simulation algorithms for 1D
  many-body quantum mechanical systems combining the Matrix Product
  State variational ansatz with Taylor, Pad\'e and Arnoldi
  approximations to the evolution operator. By comparing with previous
  techniques based on MPS and DMRG we demonstrate that the Arnoldi
  method is the best one, reaching extremely good accuracy with
  moderate resources. Finally we apply this algorithm to studying how
  correlations are transferred from the atomic to the molecular cloud
  when crossing a Feschbach resonance with two-species hard-core
  bosons in a 1D optical lattice.
\end{abstract}

\submitto{\NJP}

\maketitle

\section{Introduction}

The Density Matrix Renormalization Group (DMRG) method is a successful
technique for simulating large low-dimensional quantum mechanical
systems \cite{dmrg}. Developed for computing ground states of 1D
Hamiltonians, it is equivalent to a variational ansatz known as Matrix
Product States (MPS) \cite{ostlund95,verstraete04a}.  This relation
has been recently exploited to develop a much wider family of
algorithms for simulating quantum systems, including time evolution
\cite{vidal04,verstraete04b}, finite temperature
\cite{zwolak04,verstraete04b} and excitation spectra \cite{porras06}.
Some of these algorithms have been translated back to the DMRG
language \cite{white93,daley04,gobert05} using optimizations developed
in that field and introducing other techniques such as Runge-Kutta or
Lanczos approximations of the evolution operator
\cite{feiguin04,schmitteckert04,manmana05,manmana06}.

The MPS are a hierarchy of variational spaces, ${\cal S}_D,$ [See
Eq.~(\ref{SD})] sorted by the size of its matrices, $D.$ MPS can
efficiently represent many-body states of 1D systems, even when the
Hilbert space is so big that the coefficients of a pure state on an
arbitrary basis cannot be stored in any computer. While the accuracy
of this representation has been proven for ground
states \cite{verstraete05}, evolution of an arbitrary state changes the
entanglement among its parties, and a MPS description with moderate
resources (small $D$) might stop to be feasible.

We will take a pragmatic approach. First of all, most algorithms in
this work can compute truncation errors so that the accuracy of
simulations remains controlled. Second, we are interested in
simulating \textit{physically} small problems, such as the dynamics of
atoms and molecules in optical lattices. For such problems small $D$
are sufficient to get a qualitative and even quantitatively good
description of the observables in our systems.  As we will see below,
the biggest problem is the accumulation of truncation errors and not
always the potential accuracy of a given MPS space for representing
our states.

The outline of the paper is as follows. In Sect.~\ref{sec:mps} we
briefly introduce MPS and review some of their properties. In
Subsect.~\ref{sec:projection} we present the optimal projection onto a
MPS space, which is the keystone of most evolution algorithms. In
Sect.~\ref{sec:dmrg} we introduce for completeness the DMRG algorithm,
focusing on the concepts which are essential for time evolution. In
particular, we concentrate on the difficulties faced when implementing
DMRG simulations and how those techniques relate to MPS. In
Sect.~\ref{sec:time-evolution} we review almost all recently developed
simulation algorithms under the common formalism based on the optimal
truncation operator. Additionally we introduce three new methods: two
of them are based on Taylor and Pad\'e expansions of the evolution
operator while the other one uses an ``Arnoldi'' basis of MPS increase
the accuracy. Sect.~\ref{sec:comparison} is a detailed comparison of
MPS and DMRG algorithms using spin$-\frac{1}{2}$ models. Our study
shows that all methods are strongly limited by truncation and rounding
errors. However, among all techniques, MPS methods and in particular
our Arnoldi method performs best for fixed resources, measured by the
size of matrices $D$ or size of basis in DMRG. In the last part of our
paper, Sect.~\ref{sec:atoms} we present a real-world application of
the Arnolid evolution algorithm, which is to study a model of
hard-core bosonic atoms going through a Feschbach resonance.  Current
experiments \cite{thalhammer06,stoferle06,volz06} with such systems
have focused on the number and stability of the formed molecules. In
this work we focus on the 1D many-body states and show that coherence
is transferred from the atomic component to the molecular one, so that
this procedure can be used to probe higher order correlations in the
atomic cloud. Finally, in Sect.~\ref{sec:conclusions} we summarize our
results and open lines of research.

%%%%%%%%%%%%%%%%%%%%%%%%%%%%%%%%%%%%%%%%%%%%%%%%%%%%%%%%%%%%%%%%%%%%%%

\section{Matrix Product States (MPS)}
\label{sec:mps}

In this first section we introduce the notion of Matrix Product State,
together with some useful properties and an important operator, the
projection onto a MPS space. This section is a brief review of the
concepts found in Refs. \cite{verstraete04a,verstraete04b}.

\subsection{Variational space}

The techniques in this work are designed for the study of
one-dimensional or quasi-one-dimensional quantum mechanical lattice
models. If $N$ is the number of lattice components and $d$ the number
of degrees of freedom of each site, the Hilbert space of states will
have a tensor product structure, ${\cal H} = {\cal H}_1^{\otimes N},$
where $d=\mathrm{dim}\,{\cal H}_1$ are the degrees of freedom of a
single site. We will consider two examples here: one with spin-$1/2$
particles, where $d=2$ (Sect.~\ref{sec:comparison}), and later on a
study of atoms and molecules in an optical lattice where $d=25$
(Sect.~\ref{sec:atoms}).

Given those prerequisites, the space of MPS of size $D$ is defined as
the set of states of the form
\begin{equation}
  \label{SD}
  {\cal S}_D := \left\{
    \left(\Tr\prod_k A_k^{s_k}\right)|s_1\ldots s_N\rangle,~
    A_k^{s_n} \in \mathbb{C}^{D_k\times D_{k+1}}\right\},
\end{equation}
where $s_k=1\ldots d$ labels the physical state of the $k$-th lattice
site. The $A_k$ are complex matrices of dimensions that may change
from site to site but are of bounded size, $D_{k-1}\times D_k \leq
D^2.$ Throughout the paper we will use different notation for the
matrices $\{A_k^{s_k}\}.$ An index $k$ or $l$ will always label the
site they belong to and, whenever the expression is not ambiguous, the
site index will be dropped: $A^{s_k}:= A_k^{s_k}.$ The MPS components
can also be regarded as tensors, $A_{\alpha\beta}^{s_k},$ the Greek
indices denoting virtual dimensions $\alpha,\beta = 1\ldots D.$
Finally, at some point we will rearrange all values forming a vector
$\vec{A}_k^{t}:=(A_{11}^1,A_{11}^2,\ldots, A_{DD}^d)$ in a complex
space $\mathbb{C}^{d\times D\times D}.$

The first important property of the MPS is that they do not form a
vector space. In general, a linear combination of $M$ states with size
$D$ requires matrices of size $MD$ to be represented. It is easy to do
a constructive proof of the previous fact, but the reader may convince
himself with a simple example, made of the two product states
$|0\rangle^{\otimes N}$ and $|1\rangle^{\otimes N},$ which live in
${\cal S}_1,$ and the GHZ state $|0\rangle^{\otimes N} +
|1\rangle^{\otimes N}\in S_2.$

The previous remark leads us to another property, which is that the
dimension $D$ required to represent a state faithfully\footnote{By
  this we mean with absolutely no error.} is related to the amount of
entanglement in the system. More precisely, that dimension is equal to
the maximum Schmidt number of the state with respect to any
bipartition and thus an entanglement monotone \cite{vidal03}. Indeed,
creating entanglement forces us to use bigger and bigger MPS and this
is the reason why some problems cannot be simulated efficiently using
MPS.

The third important property is that we can efficiently compute scalar
products, distances and in general expectation values of the form
$\langle \psi|O_1 \otimes O_2 \otimes \cdots \otimes O_L|\phi\rangle,$
where $\psi,\phi\in{\cal S}_D$ and $\{O_i\}_{i=1}^L$ are local
operators acting on the individiual qubits or components of our tensor
product Hilbert space. For instance, given that we know the matrices
of those states, $\{A^{s_k}_k\}$ for $\psi$ and $\{B^{s_k}_k\}$ for
$\phi,$ the previous expectation value is made of a product of $N$
matrices,
\begin{equation}
  \langle \psi|\otimes_{k=1}^L O_k|\phi\rangle =
  \Tr \left[\prod_{k=1}^L E(O_k,A_k,B_k)\right],
  \label{expected-value}
\end{equation}
where the ``transfer matrices'' are defined as follows
\begin{equation}
  E(O_k,A_k,B_k) := \sum_{i} (A_k^{s_k})^\star \otimes B_k^{s_k}
  \langle s_k |O_k|s_k\rangle.
  \label{transfer}
\end{equation}
Since all usual Hamiltonians and correlators can be decomposed as sum
of products of local terms, the previous formulas are very useful. An
important remark is that when computing Eq.~(\ref{transfer}) one should
not directly multiply the matrices $E$, but cleverly contract the $A$
and $B$ tensors alternatively, so as to achieve a performance
${\cal O}(dD^3)$.

The last property is that expectation values, distances
$\Vert\psi-\phi\Vert^2,$ fidelities $|\langle\psi|\phi\rangle|$ and
norms $\Vert\psi\Vert^2,$ are quadratic forms with respect to each of
the matrices in the MPS.  Regarding the matrices of the state as
elements of a complex vector, we can rewrite
Eq.~(\ref{expected-value}) for the $k$-th site as
\begin{equation}
  \label{quadratic-1}
  \langle \psi|\otimes_{k=1}^L O_k|\phi\rangle = \vec{A}_k^\dagger Q \vec{B}_k,
\end{equation}
where the quadratic form $Q$ is built as follows
\begin{equation}
  \label{quadratic-2}
  Q_{[s\alpha\alpha'][s'\beta\beta']} :=   \langle s|O_k|s'\rangle
\left[\prod_{j=\{k+1\ldots L,1\ldots k-1\}} E(O_{j},A_{j},B_{j})
  \right]_{\alpha\beta,\alpha'\beta'},
\end{equation}
where $[s\alpha\beta]$ denotes grouping of indices consistent with the
ordering of the elements of $\vec{A}$ and $\vec{B}$.  This formula is
used on all algorithms, from the computation of ground states
\cite{verstraete04a} to the time evolution \cite{verstraete04b}. An
important optimization is to avoid computing the full matrix $Q$, but
to use the structure in (\ref{quadratic-2}) together with the sparsity
of the transfer matrices.

\subsection{Projector operator}
\label{sec:projection}

Even if the MPS do not form a vector space, they are embedded in a
bigger Hilbert space and provided with a distance. It is therefore
feasible, given an arbitrary state vector, to ask for the best
approximation in terms of MPS of a fixed size. The optimal projection
onto ${\cal S}_D$ is a highly nonlinear operation and it will be
denoted in this work by ${\cal P}_D.$ Following
Ref.~\cite{verstraete04b}
\begin{equation}
  \label{PDk}
  {\cal P}_{D}\sum_k c_k |\phi^{(k)}\rangle
  :=\underset{\psi\in{\cal S}_{D}}{\argmin}
  \left\Vert |\psi\rangle - \sum_k c_k|\phi^{(k)}\rangle\right\Vert^2
\end{equation}
If we rather want to approximate the action of an operator that can be
decomposed as $U=X^{-1} Y,$ we will apply a generalization of the
correction vector method \cite{dmrg}
\begin{equation}
  \label{PD}
  {\cal P}_{D}\left(X^{-1}Y | \phi \rangle \right)
  := \underset{\psi \in S_{D}}{\argmin}
  \Vert X|\psi\rangle - Y|\phi\rangle\Vert.
\end{equation}
This formula is simple to apply and behaves well for a singular
operator $X.$ Its use will become evident when studying
time evolution algorithms in Sect.~\ref{sec:time-evolution}.

One may quickly devise a procedure for computing the
minimum of Eq.~(\ref{PD}) based on the definition of distance:
\begin{eqnarray}
  \Vert X|\psi\rangle - Y|\phi\rangle\Vert^2
  =\langle\psi|X^\dagger X |\psi\rangle
  -2\Re\langle\psi|X^\dagger Y|\phi\rangle
  + \langle\phi|Y^\dagger Y |\phi\rangle.\label{distance}
\end{eqnarray}
All scalar products in Eq.~(\ref{distance}) are quadratic forms with
respect to each of the matrices in the states $|\phi\rangle$ and
$|\psi\rangle.$ The distance is minimized by optimizing these
quadratic forms site by site, or two sites at a time\footnote{The
  two-sites alternative, borrowed from DMRG, has the advantage of not
  getting trapped in subspaces of conserved quantities that conmute
  with both $X$ and $Y.$}, sweeping over all lattice sites until
convergence to a small value which will be the truncation error
\cite{verstraete04b}.

In short the algorithm looks as follows: (i) Compute some initial
guess for the matrices of the optimized state $\psi.$ (ii) Focus on
the site $k=1.$ (iii) Use Eq.~(\ref{quadratic-1})-(\ref{quadratic-2})
to find out the quadratic form associated to Eq.~(\ref{distance}) for
the first matrix
\begin{equation}
  \epsilon :=
  \vec{A}_k^\dagger Q_{X^\dagger X} \vec{A}_k - 2\Re \vec{A}_k^\dagger
  Q_{X^\dagger Y} \vec{B}_k + \vec{B}_k^\dagger Q_{Y^\dagger Y}\vec{B}_k.
  \label{error}
\end{equation}
(iv) The stationary points of the error are given by equation
\begin{equation}
  Q_{X^\dagger X} \vec{A}_k = Q_{X^\dagger Y}\vec{B}_k.
  \label{optimization}
\end{equation}
Solve this equation and use the outcome as the new value of
$A_k.$ (v) Estimate the error using Eq.~(\ref{error}).
If $\epsilon$ is small enough or does not improve significantly,
stop. Otherwise move to another site, $k=k\pm 1,$ and continue on
step (iii).

%%%%%%%%%%%%%%%%%%%%%%%%%%%%%%%%%%%%%%%%%%%%%%%%%%%%%%%%%%%%%%%%%%%%%%

\section{MPS and Density Matrix Renormalization
Group (DMRG)}
\label{sec:dmrg}

Even though the numerical techniques for dealing with MPS seem very
different from those of DMRG \cite{dmrg}, both methods are intimately
connected. First of all, the DMRG produces MPS at each state of its
algorithms. Second, particular algorithms such as the search for
ground states in open boundary condition problems are equivalent in
DMRG and MPS \cite{verstraete04a}. Third, other concepts, such as
basis adaptation and state transformation from DMRG are analogous to
the MPS ones, even though they are less powerful and less accurate. In
this section we will elaborate on these statements.

\subsection{DMRG builds matrix product states}

\begin{figure}[t]
  \includegraphics[width=0.5\linewidth]{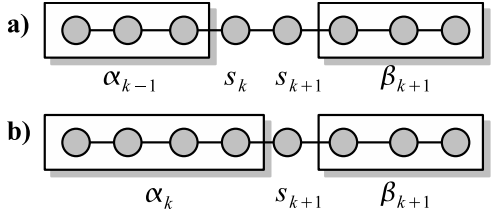}%
  \caption{(a) DMRG view of a state, with the basis for the left
    $\ket{\alpha_{k-1}}$ and right block $\ket{\alpha_{k+1}},$ and the
    states of the central spins, $\ket{s_k}, \ket{s_{k+1}}.$ (b) On a
    renormalization step, one spin is incorporated to the left block
    and a new basis is built, $\ket{\alpha_{k}} := A_{k-1,k}^{s_k}
    \ket{\alpha_{k-1}}\ket{s_k}.$}
  \label{fig-dmrg}
\end{figure}

The DMRG algorithms are based on the key idea that interesting states
can be expressed using a basis with a small number of vectors. Take
for instance the chain in Fig.~\ref{fig-dmrg}a. Any state of this
chain can be decomposed in the form
\begin{equation}
  \label{dmrg-state}
  \ket{\psi} = \psi(\alpha_{k-1}s_ks_{k+1}\alpha_{k+1})
  \ket{\alpha_{k-1}}\ket{s_k}\ket{s_{k+1}}\ket{\alpha_{k+1}},
\end{equation}
where we sum over repeated indices. While for describing individual
spins we use all possible states, $s_{k},s_{k+1}=1\ldots d,$ in DMRG the
states of the left and right blocks are expressed in a finite basis,
$\alpha_k,\beta_k = 1\ldots M,$ where $M,$ the number of states kept,
is the basic control parameter. DMRG algorithms build those basis
states recursively, by taking a smaller block, adding a site and
truncating the basis of the bigger block ``optimally'' in a way ot be
precised later. Thus we have the relations \numparts
  \begin{eqnarray}
  \ket{\alpha_k} := A_{\alpha_{k-1}\alpha_{k}}^{s_k}
  \ket{\alpha_{k-1}}\ket{s_{k-1}},\label{basis-left}\\
  \ket{\beta_k} := A_{\beta_k\beta_{k+1}}^{s_{k+1}}
  \ket{s_{k+1}}\ket{\beta_{k+1}}.\label{basis-right}
  \end{eqnarray}
\endnumparts
It is trivial to see, substituting those equations into
Eq.~(\ref{dmrg-state}) that all DMRG states are matrix product states
\cite{ostlund95,verstraete04a}.

%----------------------------------------------------------------------

\subsection{Targetting}
\label{sec:targetting}

An important question in DMRG is how many states we have to keep for
the left and right blocks and how to optimize them. The criterium is
that the basis describing those blocks has to represent accurately a
family of target states, $\ket{\phi_n}.$ The algorithm consists on a
series of sweeps over the lattice
\cite{feiguin04,manmana05,manmana06,schollwoeck06} with a recipe to
achieve an optimal representation. For instance, let us say that we
have an approximate basis around sites $k$ and $k+1$ and we will
improve this sweeping from left to right, to $k+1$ and $k+2$
[Fig.~\ref{fig-dmrg}]. The first step is to build the weighted density
matrix of the left piece of the chain
\begin{eqnarray}
\fl \rho_L := \sum_n w_n \phi_n(\alpha_{k-1}s_ks_{k+1}\beta_{k+1})
  \phi_n(\alpha_{k-1}'s_k's_{k+1}\beta_{k+1})^\star
  \ket{\alpha_{k-1}s_k}\bra{\alpha_{k-1}'s_k'}\nonumber\\
  =\sum_n w_n \Tr_{s_{k+1}\alpha_{k+1}}
  \ket{\phi_n}\bra{\phi_n},  \label{rho}
\end{eqnarray}
with some normalized weights, $\sum_n w_n=1.$ Second, take the $M$
most significant eigenvectors of this matrix
\begin{equation}
  \rho\ket{\alpha_k} = \lambda_k\ket{\alpha_k},\quad
  \lambda_k \geq \lambda_j\,\quad \forall k>l;\;k,l=1\ldots M\times d.
\end{equation}
These vectors become the improved new basis for the enlarged left
block and are related by a transformation matrix to the basis elements
of the smaller block (\ref{basis-left}). Matrix
elements of observables and of the initial and target states have to
be recomputed using this isometry and one continues until the end of
the lattice. A similar procedure is employed to create the vectors
$\ket{\beta_k}$ recursively from right-to-left. Multiple sweeps can be
performed this way.

For static problems one uses as target states, $\ket{\phi_n},$ the
ground state and a number of excitations, computed with the
Hamiltonian in the truncated basis. In some time evolution algorithms
\cite{feiguin04,schmitteckert04,manmana05,manmana06,schollwoeck06} the
targetting is done with respect to approximations of the time evolved
state $\ket{\psi(t)}.$ In this work we have used $\ket{\phi_k} :=
\ket{\psi(k\Delta t/(N_v-1))},$ where the intermediate and final
states, $\ket{\psi(\tau)}$ are computed using the Hamiltonian on the
truncated basis [See Sect.~\ref{sec:rk}].

\subsection{Targetting vs. projection}
\label{sec:mps-vs-dmrg}

\begin{figure}[t]
\centering
  \includegraphics[width=0.5\linewidth]{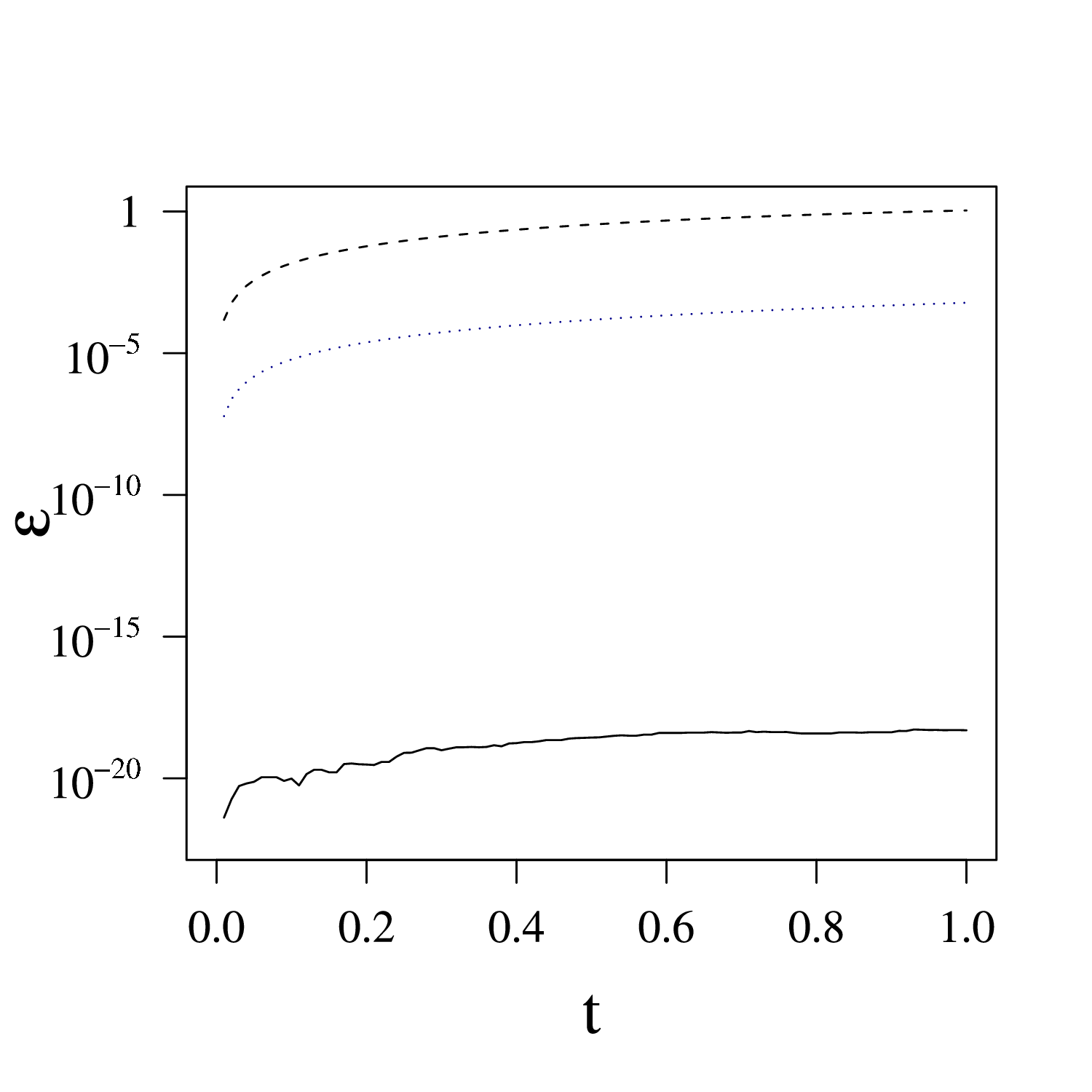}%
\caption{Error in the DMRG Runge-Kutta algorithm [See
  Sect.~\ref{sec:rk}], using as initial state
  $\ket{\psi}=\ket{0}^{\otimes N}$ and Hamiltonians
  $H=\sum_i\sigma^x_i$ (solid, dashed) or $H=\prod_i \sigma^x_i$
  (dotted). The algorithm used either a single basis (dashed) or two
  set of basis states (solid, dotted) for targetting. The dashed line
  shows errors due to using a single basis, the dotted line fails
  because the Hamiltonian does not have nonzero elements on the
  initial DMRG basis.}
  \label{fig-targetting}
\end{figure}

There are a number of complicated subtleties and facts that are rarely
mentioned in the DMRG literature. The first one is that in most
implementations of the targetting algorithm, the state itself is
updated and rewritten in the new basis after each step
\begin{equation}
  \ket{\psi} \to
  \psi(\alpha_{k-1}s_ks_{k+1}\beta_{k+1})
  A^{s_k\star}_{\alpha_{k-1}\alpha_k}
  \ket{\alpha_k}\ket{s_k}\ket{s_{k+1}}\ket{\beta_{k+2}}.
\end{equation}
However, this leads to wrong answers because the initial state
$\psi(0)$ deteriorates during the algorithm, and it only works when
the basis of the left and right block are already large enough. A more
accurate procedure consists on keeping the initial state in the
original basis, and on each step of the the algorithm compute its
matrix elements in the new basis.  As an example, in
Fig.~\ref{fig-targetting} we plot the error of an evolution algorithm
with the trivial algorithm and with the correct one, for an initial
product state $\ket{1}^{\otimes L}$ in the $\sigma^z_i$ basis and
using a simple Hamiltonian $H=\sum_i \sigma^x_i.$ In both cases we
begin with a basis of size $M=1$ vectors per block, letting the basis
grow up to $2^{L/2}.$ However only the second method is capable of
doing the update.

An even more important issue is that if in current literature this
method is formulated in an intuitive way and little is known about why
it works and how accurate it is. To this respect, a close look at the
matrices (\ref{basis-left})-(\ref{basis-right}) created by the
targetting with multiple vectors, $\ket{\phi_t},$ shows that their
procedure is quite similar to the projection operator for some linear
combination ${\cal P}_M (\sum_k c_k \ket{\phi_k}),$ but the steps for
computing the optimal approximation are wrong if there are more than
one target state.

Another consequence of being tied to the notion of ``optimal basis''
instead of treating MPS as a variational family of wavefunctions, is
that the targetting procedure does not work when the operators $O_n$
and the target states $\ket{\phi_n}$ have no elements in the initial
basis. A trivial example consits on the same product state
$\ket{\psi(0)} = \ket{1}^{\otimes N}$ as before, beginning with $M=1$
states per block and evolved with $H=\prod_i \sigma^x_i.$ As shown in
Fig.~\ref{fig-targetting}, a conventional DMRG sweep cannot enlarge
the basis in the appropiate way and the method fails to follow the
evolution of the state. A MPS algorithm, on the other hand, even if it
begins with a state with $D=1$, grows up the state up to the maximal
size $D=2$ and makes absolutely no error in the simulation.

The final practical difference is that in current DMRG works
\cite{feiguin04,schmitteckert04,manmana05,manmana06} the powers of the
Hamiltonian acting on the original state, $H^n\ket{\phi},$ are
approximated by the truncation of $H$ to the current basis. This is
another potential source of errors which we cleverly avoid in our
Arnoldi and Runge-Kutta methods shown below [Sect.~\ref{sec:rkmps} and
\ref{sec:arnoldi}] and its effect are be evident in some of the
simulations [Sect.~\ref{sec:comparison}].

The previous paragraphs can be rephrased as follows: DMRG algorithms
need not only an initial input state, but also a suitable and large
enough basis for doing the time evolution. How to construct this basis
is largely heuristics, and contrasts with the systematic way in which
MPS work.

%%%%%%%%%%%%%%%%%%%%%%%%%%%%%%%%%%%%%%%%%%%%%%%%%%%%%%%%%%%%%%%%%%%%%%

\section{Time evolution}
\label{sec:time-evolution}

Since ${\cal S}_D$ is not a vector space, we cannot solve a
Schr\"odinger equation directly on it. Our goal is rather to
approximate the evolution at short times by a formula like
\begin{equation}
  |\psi(t+\Delta t)\rangle \simeq {\cal P}_D \left[U_n(\Delta t)
  |\psi(t)\rangle\right].
\label{evolution}
\end{equation}
Here, ${\cal P}_D$ is the optimal projection operator defined before
and $U_n(\Delta t) = \exp(-i H \Delta t) + {\cal O}(\Delta t^n)$ is
itself an approximation of $n$-th order to the evolution operator.
Even though this formulation applies qualitatively to all recently
developed MPS and DMRG algorithms, there are subtle differences that
make some methods more accurate than other. The actual implementation
of time evolution is thus the topic of the following subsection.

\subsection{Trotter decomposition}

While there had been previous works on simulating time evolution using
DMRG algorithms \cite{cazalilla02,cazalilla03,luo03}, an important
breakthrough happened with the techniques in Ref.~\cite{vidal04},
which was later on translated to the DMRG language
\cite{white93,daley04,gobert05} and have since been applied to a
number of interesting problems
\cite{winkler06,kollath05,micheli04,clark04}.  Vidal's seminal paper
suggested doing the time evolution with a mixture of two-qubit quantum
gates and truncation operations. The idea is to split a Hamiltonian
with nearest-neighbor interactions into terms acting on even and odd
lattice edges
\begin{equation}
  H = \sum_{k=1}^{L/2} H_{2k} + \sum_{k=1}^{L/2} H_{2k-1}
  =: \Heven + \Hodd.
\end{equation}
This leads to a Suzuki-Trotter decomposition of the time evolution
operator into a sequence of unitaries acting on even and odd lattice
bonds:
\begin{equation}
  e^{-iH\Delta t} \simeq
  e^{-i \Heven \Delta t}
  e^{-i \Hodd \Delta t}=  \prod_{k=1}^{L/2} e^{-iH_{2k}\Delta t}
  \prod_{k=1}^{L/2} e^{-iH_{2k+1}\Delta t}.
  \label{trotter-2}
\end{equation}
Inserting optimal truncations in between the applications of these
unitaries, $U_j := \exp(-iH_j\Delta t),$ Vidal's algorithm can be
recasted in the form of Eq.~(\ref{evolution})
\begin{equation}
  |\psi(t+\Delta t)\rangle :=
  \prod_{k=1}^{L/2} {\cal P}_D U_{2k}
  \prod_{k=1}^{L/2} {\cal P}_D U_{2k+1}
  |\psi(t)\rangle.
\end{equation}
Applying each of the unitaries $U_{j}$ is a relatively costless task,
and in particular, for open boundary condition problems, the
combination ${\cal P}_D U_{j}$ can be done in a couple of steps which
amount to contracting neighboring matrices and performing a singular
value decomposition \cite{vidal04,white93,daley04}.

In Ref.~\cite{verstraete04b} we developed a variant of this method
which does not insert so many truncation operators, but waits until
the end
\begin{equation}
  |\psi(t+\Delta t)\rangle :=
  {\cal P}_D \prod_{k=1}^{L/2} U_{2k}
  \prod_{k=1}^{L/2} U_{2k+1}
  |\psi(t)\rangle.
\end{equation}
This procedure has a similar computational cost, ${\cal O}(ND^3),$ but
the solution is then expected to be optimal for a given Trotter
decomposition and can be generalized to problems with periodic
boundary conditions.

The accuracy of either method may be increased by an order of
magnitude using a different second order decomposition
\begin{equation}
  e^{-iH\Delta t} =
  e^{-i \Heven \Delta t/2}
  e^{-i \Hodd \Delta t}
  e^{-i \Heven \Delta t/2}
  + {\cal O}(\Delta t^2),
  \label{trotter-2b}
\end{equation}
but it is better to apply a Forest-Ruth formula \cite{white93,omelyan02}
\begin{eqnarray}
  \fl e^{-iH\Delta t} = e^{-i\Heven \theta \Delta t/2}
  e^{-i\Hodd \theta \Delta t}  e^{-i\Heven (1- \theta) \Delta t/2}
  e^{-i\Hodd (1- 2\theta) \Delta t}\times\nonumber\\
\times
  e^{-i\Heven (1- \theta) \Delta t/2}
  e^{-i\Hodd \theta \Delta t}
  e^{-i\Heven \theta \Delta t/2},
  \label{forest-ruth}
\end{eqnarray}
with the constant $\theta=1/(2-2^{1/3})$ and which has a Trotter error
of order ${\cal O}(\Delta t^5).$ Note that better Suzuki-Trotter
formulas can be designed but they do not provide a big improvement in
the number of exponentials or accuracy \cite{omelyan02}.

\subsection{Runge-Kutta and Lanczos with DMRG}
\label{sec:rk}

After the development of the Trotter methods, in Ref.~\cite{feiguin04}
we find a new algorithm in the field of DMRG. The idea now is to use
not a Trotter formula, but a Runge-Kutta iteration:
\numparts
\begin{eqnarray}
  \ket{k_1}&:=&\Delta t\, H(t) \ket{\psi(0)},\\
  \ket{k_2}&:=&\Delta t\, H(t) [\ket{\psi(0)}+\case{1}{2}\ket{k_1}],\\
  \ket{k_3}&:=&\Delta t\, H(t) [\ket{\psi(0)}+\case{1}{2}\ket{k_2}],\\
  \ket{k_4}&:=&\Delta t\, H(t) [\ket{\psi(0)}+\ket{k_3}].
\end{eqnarray}
\endnumparts
These vectors are then used to interpolate the state at other
times\footnote{Notice the typo in Eq.~(4) in Ref.~\cite{feiguin04}.}
\numparts
\begin{eqnarray}
  \ket{\psi(\case{\Delta t}{3})} &\simeq&
  \ket{\psi(0)} + \case{1}{162}[31\ket{k_1}+14(\ket{k_2}+\ket{k_3})
  -5\ket{k_4}],\\
  \ket{\psi(2\case{\Delta t}{3})} &\simeq&
  \ket{\psi(0)} + \case{1}{81}[16\ket{k_1}+20(\ket{k_2}+\ket{k_3})
  -2\ket{k_4}],\\
  \ket{\psi(\Delta t)} &\simeq&
  \ket{\psi(0)} + \case{1}{6}[\ket{k_1}+2(\ket{k_2}+\ket{k_3})
  +\ket{k_4}].
  \label{rk1}
\end{eqnarray}
\endnumparts
These three vectors, together with $\ket{\psi(0)},$ are used to find
an optimal basis using the targetting procedure explained in
Sect.~\ref{sec:targetting}. Once the basis is fixed, the time
evolution is performed with the same formulas but smaller time step,
$\Delta t/10.$ There are variants of this technique which approximate
the time evolved state using a Lanczos method
\cite{noack05,hochbruck96} with the truncated matrix of the
Hamiltonian in the DMRG basis. The different submethods differ on
whether the preparation of the basis is done only using the final and
initial state \cite{manmana06} or multiple intermediate time steps
\cite{schmitteckert04,manmana05}.

\subsection{Runge-Kutta like method with MPS}
\label{sec:rkmps}

Our implementation of a 4-th order method with Runge-Kutta like
formulas uses several simplifications. First of all, since our
Hamiltonian is constant, a Runge-Kutta expansion becomes equivalent to
a fourth-order Taylor expansion of the exponential
\begin{eqnarray}
  \ket{\psi(\Delta t)} &=& \sum_{n=0}^{4} \frac{1}{n!} (iH\Delta t)^n
  \ket{\psi(0)} + {\cal O}(\Delta t^5)\nonumber\\
  &\simeq& (iH\Delta t - z_1)(iH\Delta t - z_1^\star)
  (iH\Delta t - z_2)(iH\Delta t - z_2^\star) \ket{\psi(0)}\nonumber\\
  &=:& Y_1 Y_2 Y_3 Y_4 \ket{\psi(0)}.
  \label{rk2}
\end{eqnarray}
Here we have rewritten the fourth order polynomial in terms of its
roots, $z_1$ and $z_2.$ Using the fact that we know an efficient
algorithm to compute ${\cal P}_D Y_i$ acting on a MPS, we can write
\begin{equation}
  \ket{\psi(\Delta t)} := \prod_{i=1}^4 {\cal P}_D Y_i \ket{\psi(0)},
\end{equation}
which is our MPS Runge-Kutta-like algorithm. There are multiple
reasons to proceed this way. On the one hand, we do not want to
approximate higher powers of the Hamiltonian using a truncated basis
[See Sect.~\ref{sec:targetting}]. On the other hand, if we treat
expand the Hamiltonian to all powers, there will be too many operators
and the complexity of the state will increase enormously. The previous
decomposition has proven to be a good compromise.

\subsection{Pade approximations}
\label{sec:pade}

Runge-Kutta and in general polynomial approximations to the
exponential are not unitary. Indeed, if we look at the eigenvalues of
the evolution operator in either Eq.~(\ref{rk1}) or Eq.~(\ref{rk2}),
we will see that they are of the form $\lambda_n = \sum_{n=0}^4
(iE_N\Delta t)^n/n!,$ where $E_n$ are the eigenvalues of the
Hamiltonian $H.$ From this equation we see that $|\lambda_n| \neq 1$
and some eigenmodes may grow exponentially, which is another way to
say that Runge-Kutta algorithms are numerically unstable.

There exist multiple implicit methods that eliminate the lack of
unitarity and produce stable approximations. They receive the name
``implicit'' because the value of the state at a later time step
$\Delta t$ is obtained by solving an equation or inverting an
operator. We will focus on Pad\'e approximations to the exponential
\begin{equation}
  \label{pade}
  U_n(\Delta t) = \frac{\sum_k \alpha_k H^k}{\sum_k \beta_k H^k},
\end{equation}
which are computed with same order polynomials in the numerator and
denominator \cite{moler03}. The lowest order method is known as the
Crank-Nicholson scheme, it arises from a second order discretization
of the Schr\"odinger equation and has the well known form
\begin{equation}
  \UCN(\Delta t) = \frac{1 - i H\Delta t/2} {1 + iH\Delta t /2}.
  \label{cn2}
\end{equation}
It is easy to verify that the eigenvalues of this operator are just
phases and that the total energy is a conserved quantity. The other
method that we have used and which we compare in this work is a fourth
order expansion
\begin{equation}
  \UPADE(\Delta t) =
  \frac{1 - i\Delta t H/2 - (\Delta t H)^2/12}{1 + i\Delta t H/2 -
    (\Delta tH)^2/12}.
  \label{cn4}
\end{equation}
Applying either $\UCN$ or $\UPADE$ on a matrix product state is
equivalent to solving the problem in Eq.~(\ref{PD}), where the
operators $X$ and $Y$ are the denominator and the numerator in the
previous quotients.

\subsection{Arnoldi method}
\label{sec:arnoldi}

The last and most important method that we present in this paper
combines many of the ideas explained before. First of all, we will use
the fact that a linear combination of MPS, such as $\sum_{k=1}^{N_v}
c_k|\phi_D^{(k)}\rangle,$ resides in a space of bigger matrix product
states, ${\cal S}_{N_vD},$ and it is thus a more accurate
representation of the evolved state than a single vector of size $D.$
In the language of DMRG, $N_v$ vectors each of size $D$ provide us
with an effective basis of size $N_vD.$ This optimistic estimate is
only possible when the vectors are indeed linearly independent. Our
choice for an optimal decomposition will be therefore a Gramm-Schmidt
orthogonalization of the Krylov subspace,
$\{\psi,H\psi,H^2\psi\ldots\},$ performed using MPS
\begin{equation}
  |\phi_{k+1}\rangle \simeq {\cal P}_D\left(H|\phi_{k}\rangle -
    \sum_{j\leq k} \frac{\langle\phi_j|H|\phi_k\rangle}{\langle\phi_j|\phi_j
\rangle}
    |\phi_j\rangle\right),
  \label{phikp1}
\end{equation}
with initial condition $|\phi_0\rangle := |\psi(0)\rangle.$ Defining
the matrices $N_{ik} := \langle \phi_j |\phi_i\rangle$ and $H_{ik} :=
\langle \phi_j | H |\phi_i\rangle$ we compute an Arnoldi estimate of
the exponential
\begin{equation}
  \label{Arnoldi}
  |\psi(\Delta t)\rangle :=
  {\cal P}_D \sum_k [e^{-i\Delta tN^{-1}H}]_{k0} |\phi_k\rangle.
\end{equation}
This algorithm involves several types of errors all of which can be
controlled. First, the error due to using only $N_v$ basis vectors is
proportional the norm of the vector $\phi_{N_v}$ as in ordinary
Lanczos algorithms \cite{noack05,hochbruck96}. Truncation errors
arising from ${\cal P}_D$ can be also computed during the numerical
simulation. Out of this errors, in our experience, the final
truncation in Eq.~(\ref{Arnoldi}) is the most critical one, since the
other ones may be compesanted by adding more and more vectors.

Finally, for completeness, in this work we have also implemented a
Lanczos method. It differs from the previous one in that the basis is
built orthogonalizing only with respect to the two previous vectors,
so that Eq.~(\ref{phikp1}) contains only three summands, as in
ordinary Lanczos iterations. One then assumes that, due to the
Hermiticity of the Hamiltonian, orthogonalitiy to the rest of the
Krylov basis is preserved. Furthermore, if this is true, the effective
matrices for $H$ and $N$ are tridiagonal and can be constructed with a
simple recurrence. This method has a potential gain of
$\cal{O}(1/N_v)$ in speed due to the simplifications in
Eq.~(\ref{phikp1}). However, as we will see later, truncation errors
spoil the orthogonality of the Lanczos vectors and makes the method
useless for small matrices.

%%%%%%%%%%%%%%%%%%%%%%%%%%%%%%%%%%%%%%%%%%%%%%%%%%%%%%%%%%%%%%%%%%%%%%

\section{Comparison}
\label{sec:comparison}

\begin{figure}[t]
\centering
  \includegraphics[width=0.5\linewidth]{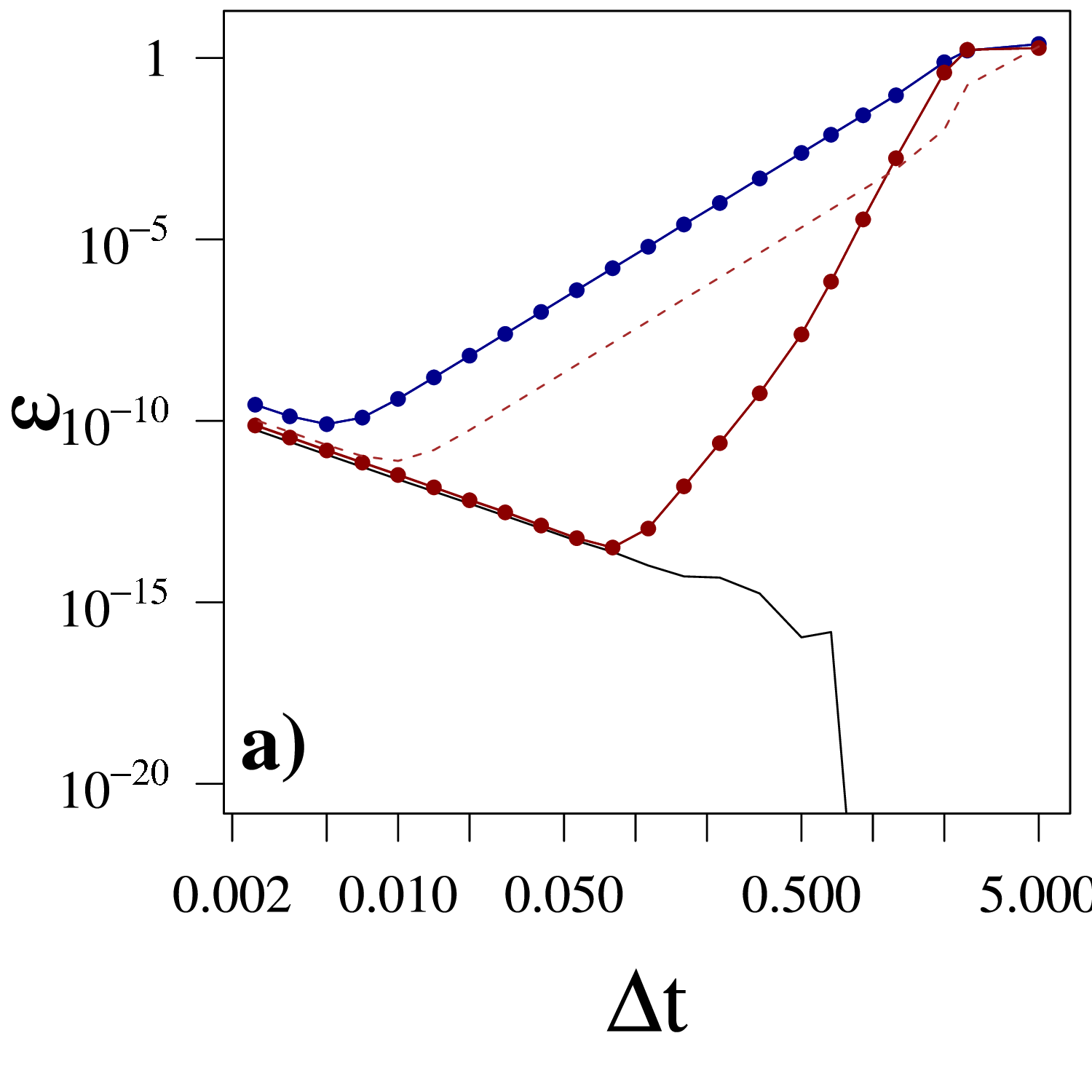}%
  \includegraphics[width=0.5\linewidth]{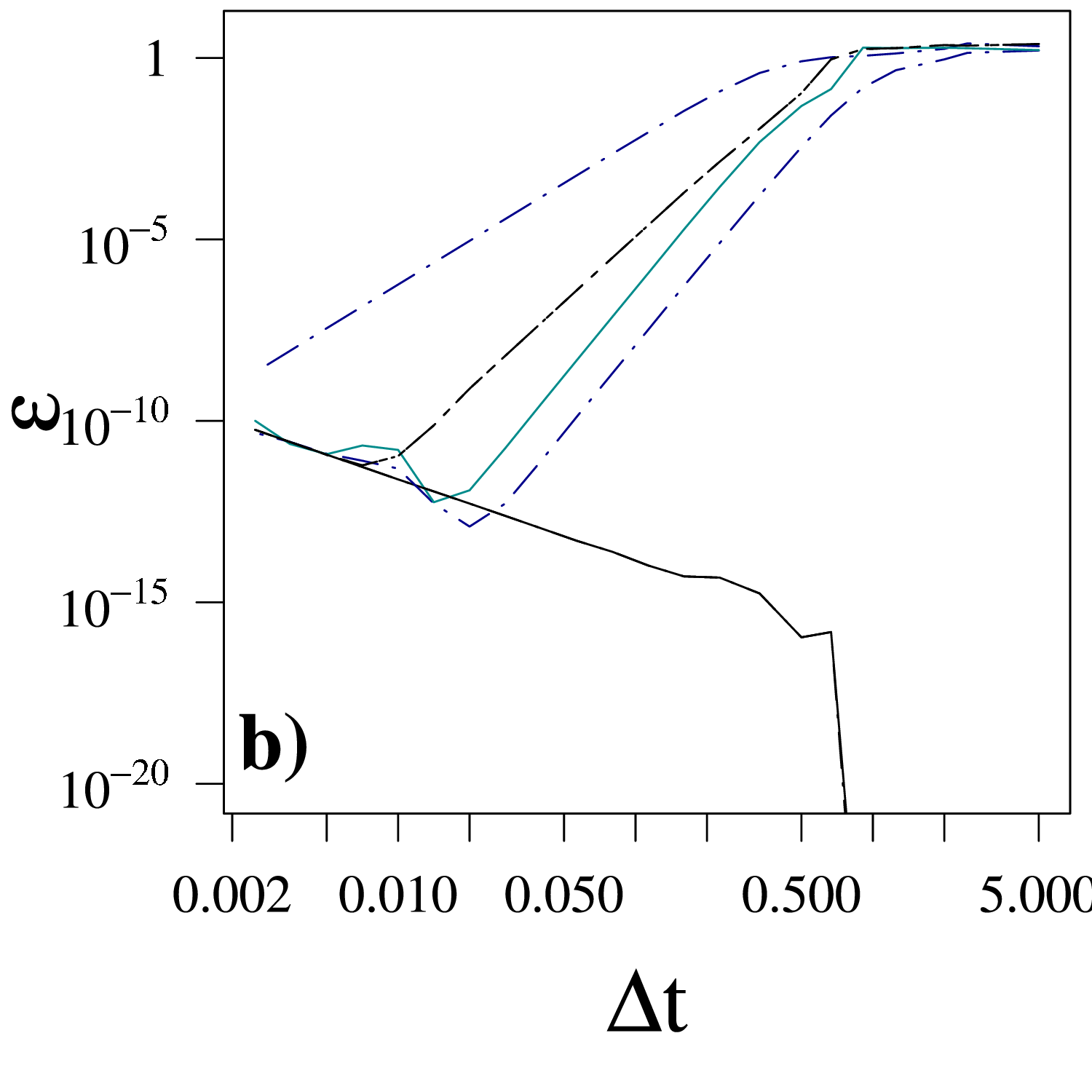}
  \includegraphics[width=0.5\linewidth]{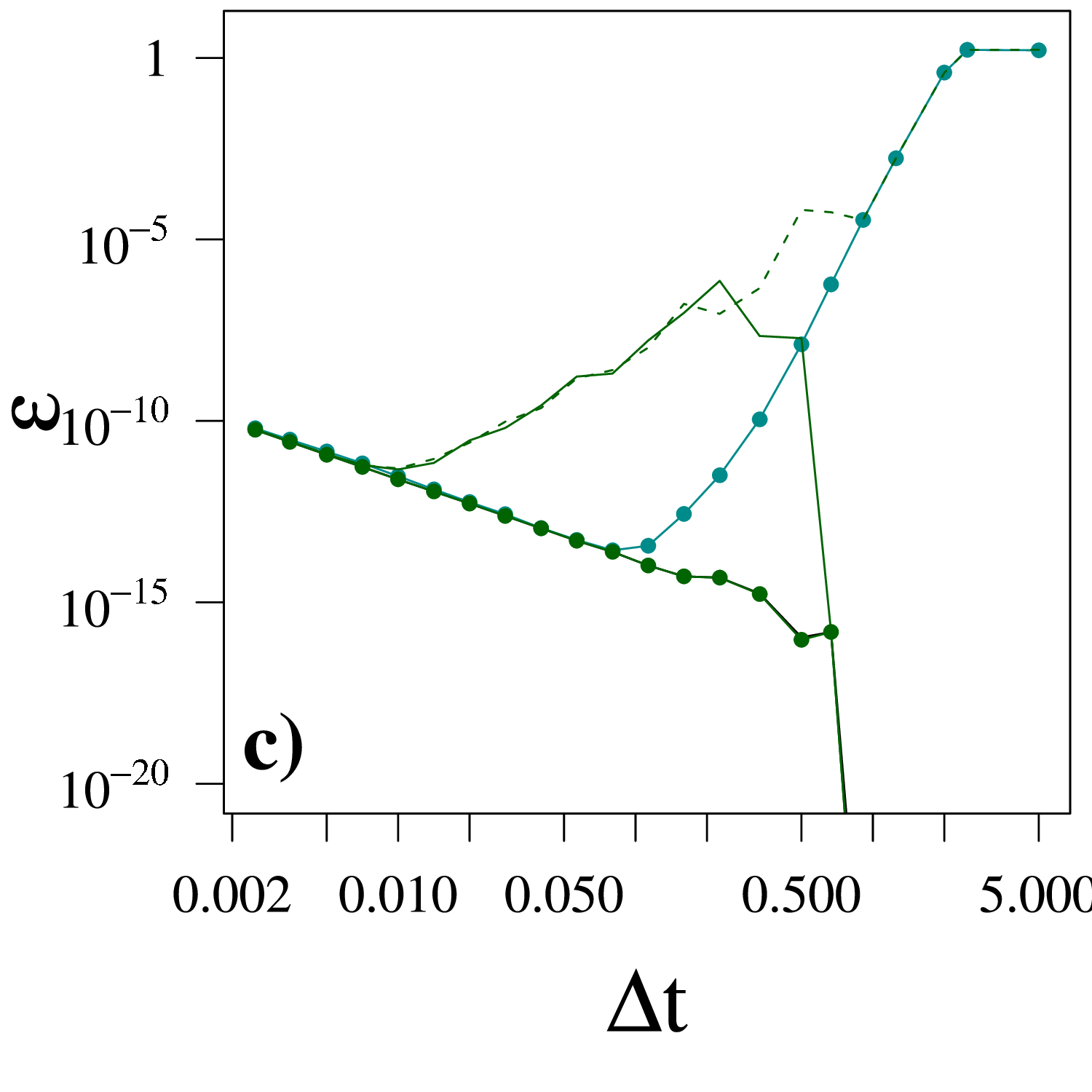}%
  \includegraphics[width=0.38\linewidth]{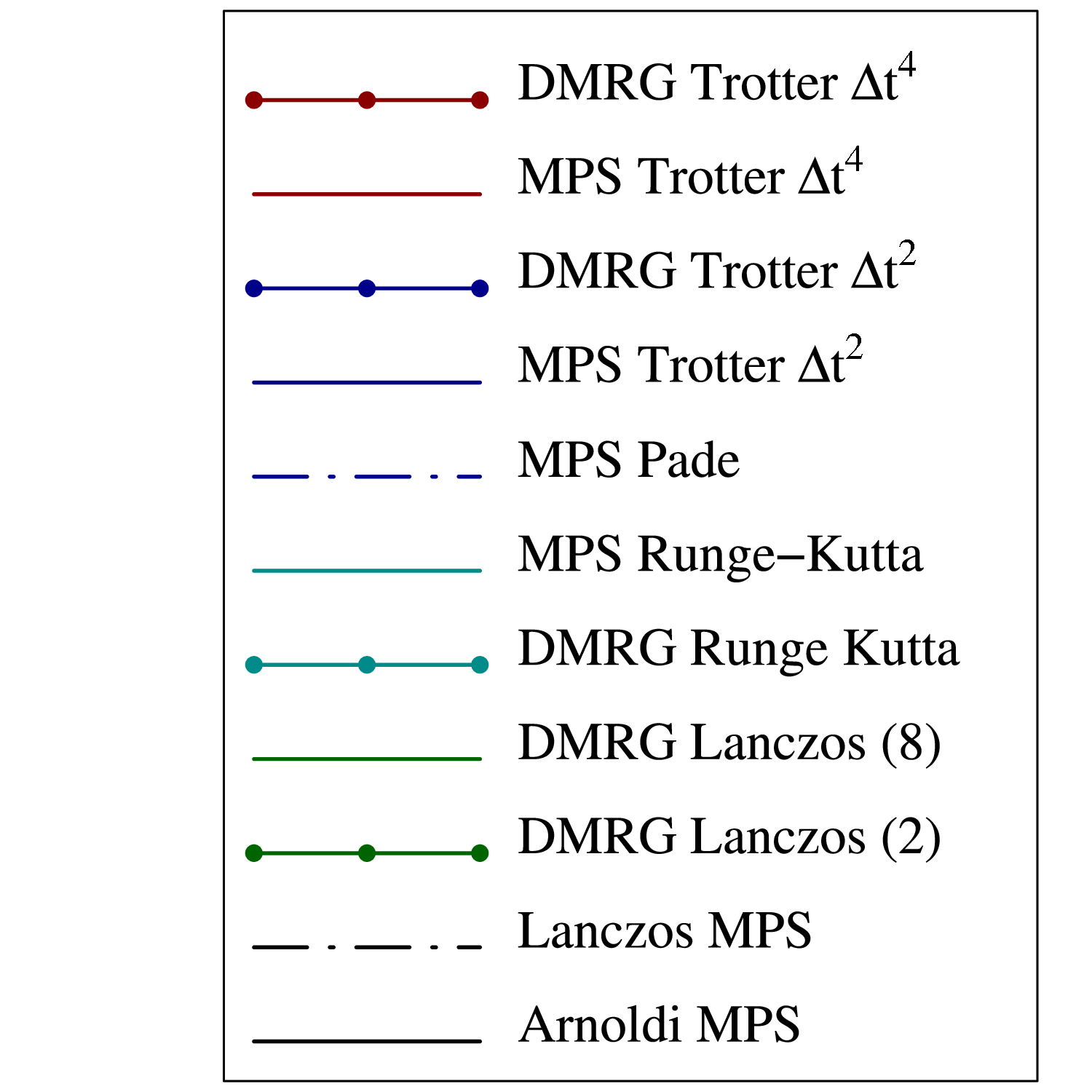}
\caption{Error, $\varepsilon,$ vs. time step, $\Delta t,$ for
  simulations of model (\ref{Hamiltonian}) with 8 spins,
  $\theta=0.35,$ $\Delta=0,$ $D=16$ and $T=10.$ We compare (a) Trotter
  methods, (b) MPS algorithms and (c) DMRG algorithms. As a reference,
  all plots contain the error of the MPS Arnoldi and Lanczos methods
  (solid black line).}
  \label{fig-exact}
\end{figure}

\begin{figure}[t]
\centering
  \includegraphics[width=0.5\linewidth]{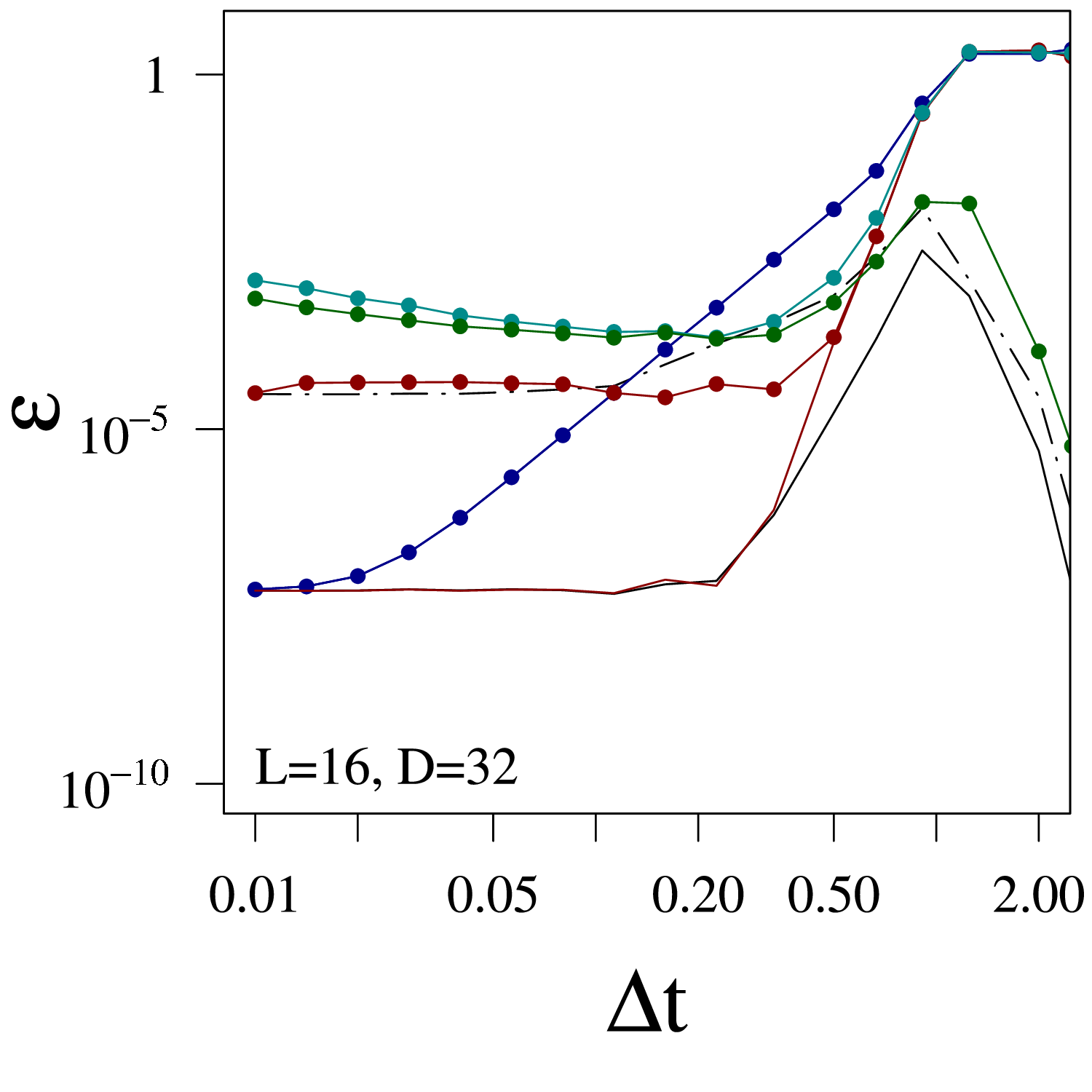}%
  \includegraphics[width=0.5\linewidth]{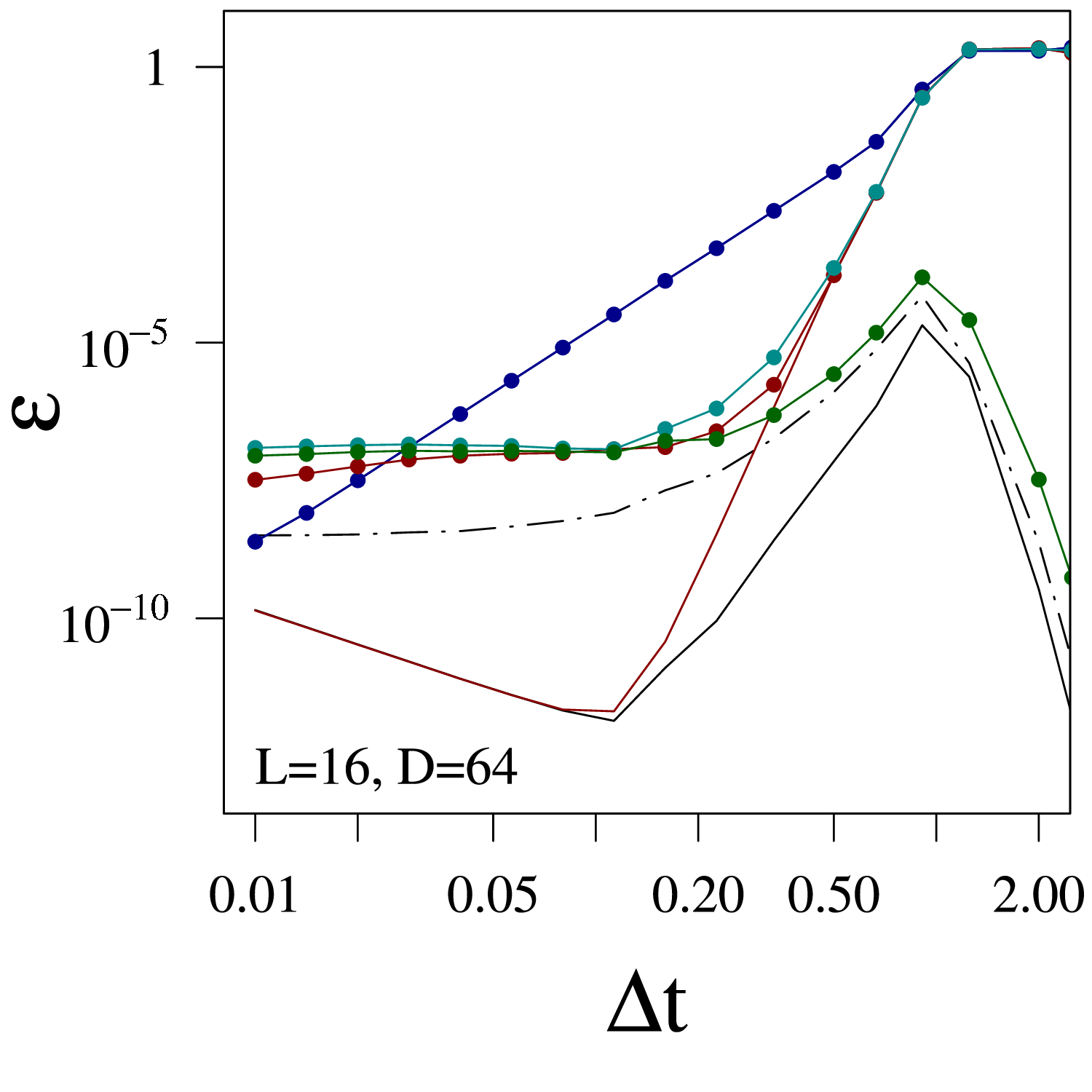}
  \includegraphics[width=0.5\linewidth]{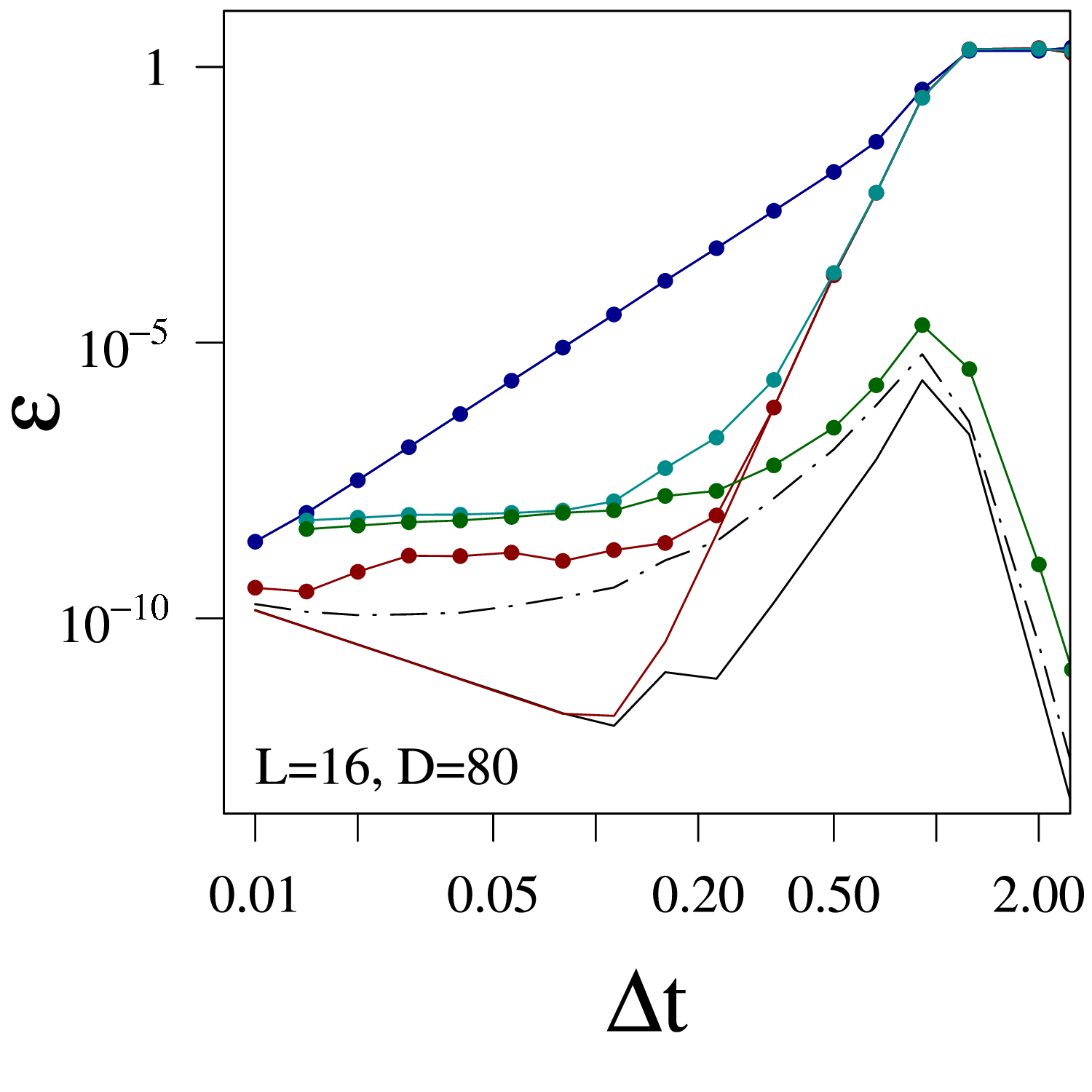}%
  \includegraphics[width=0.5\linewidth]{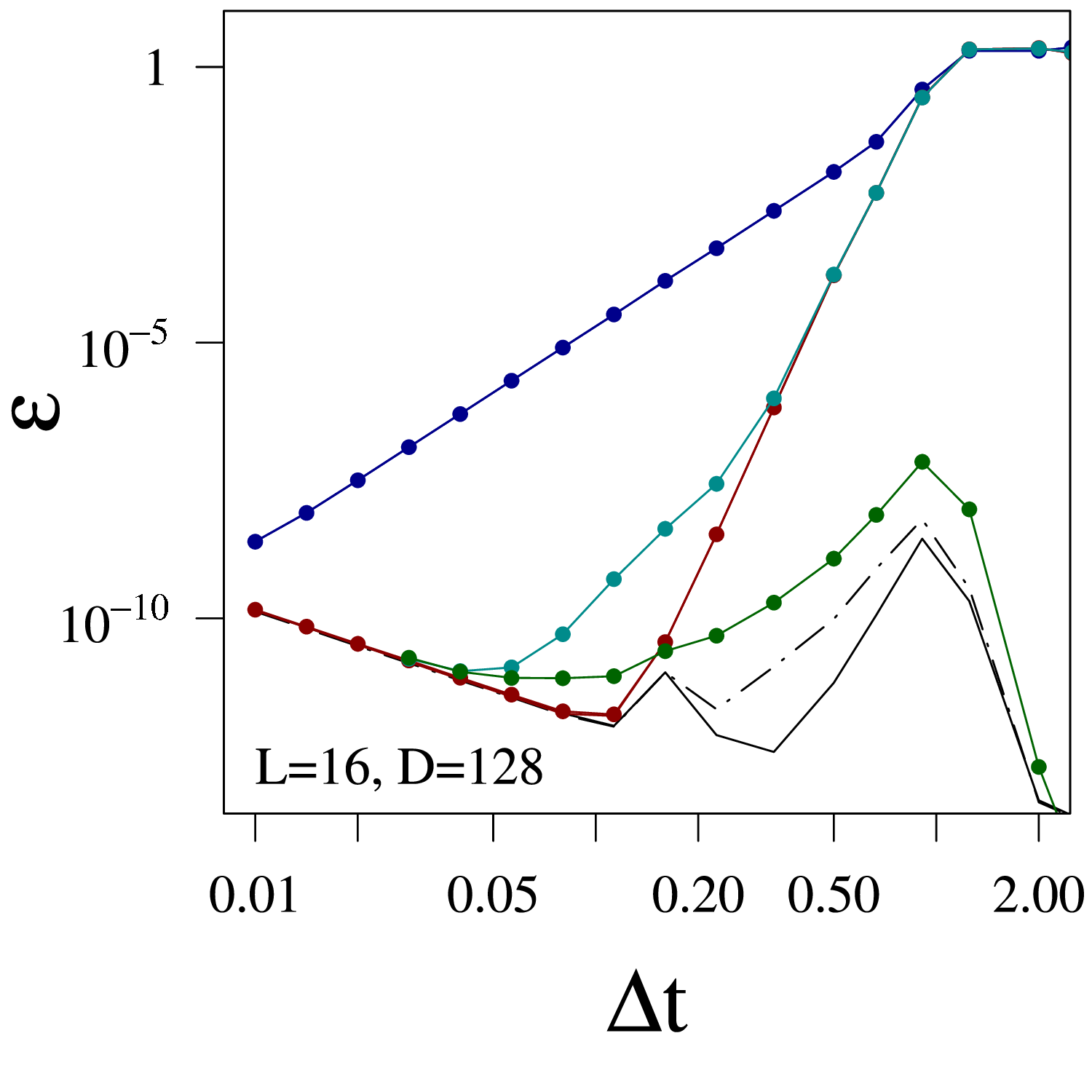}
\caption{Wavefunction error vs. time step, $\Delta t,$ for simulations
  of model (\ref{Hamiltonian}) with 16 spins, $\theta=0.35,$
  $\Delta=0,$ and $T=10.$ In Fig.~(a)-(d) we plot the outcome for
  different MPS sizes or DMRG basis, denoted by $D$. The association
  between methods and line types is that of Fig.~\ref{fig-exact}.}
  \label{fig-truncated}
\end{figure}

\begin{figure}[t]
\centering
  \includegraphics[width=0.5\linewidth]{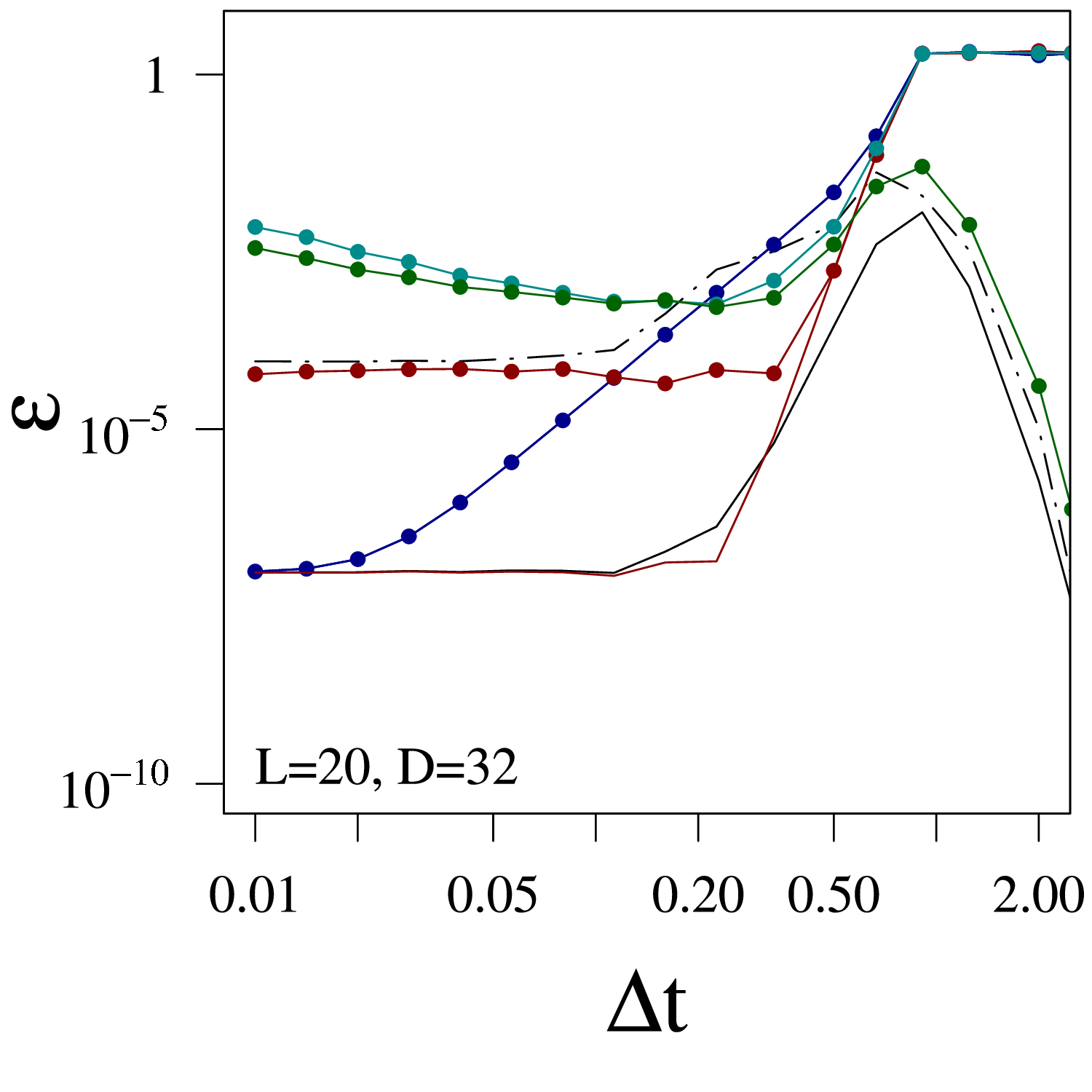}%
  \includegraphics[width=0.5\linewidth]{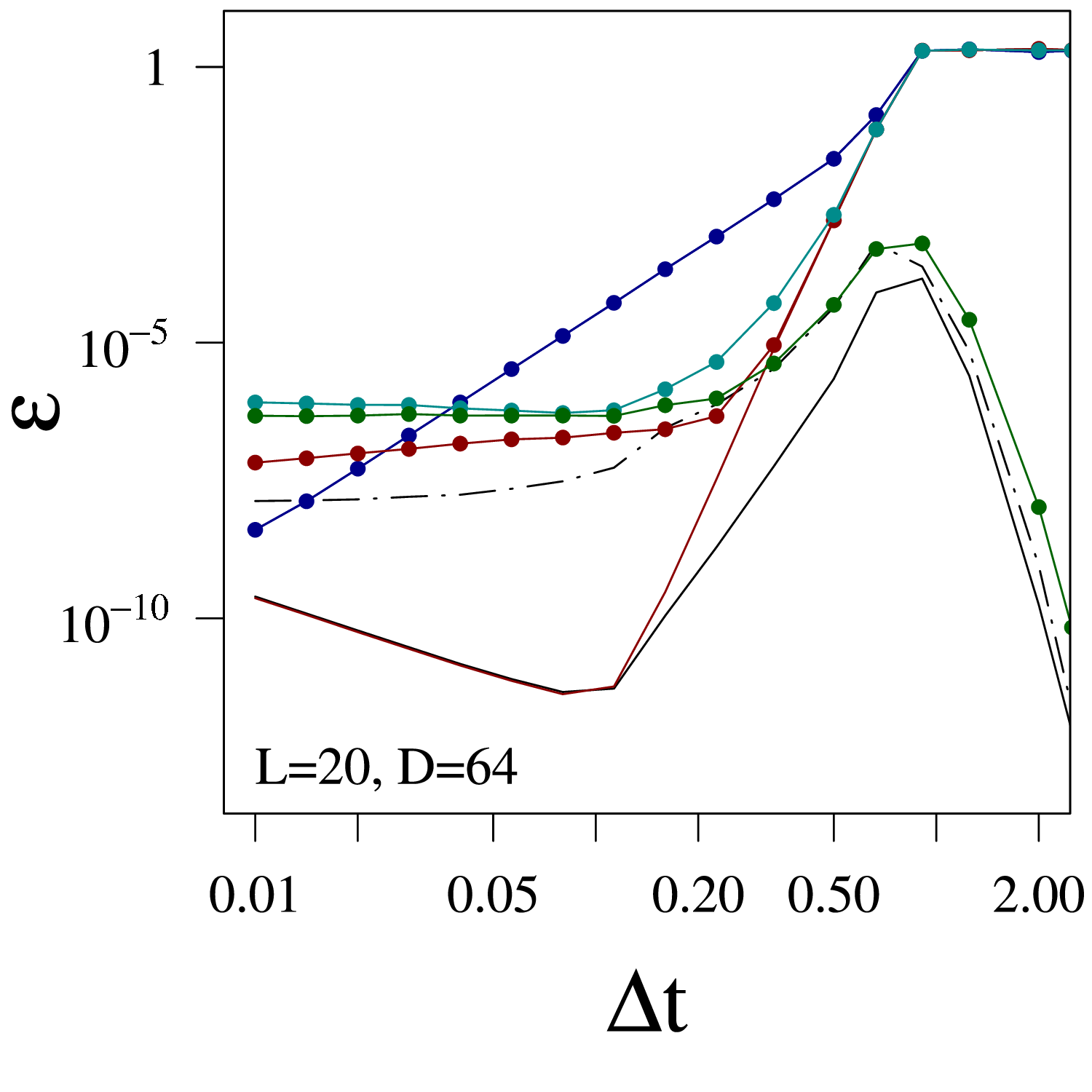}
  \includegraphics[width=0.5\linewidth]{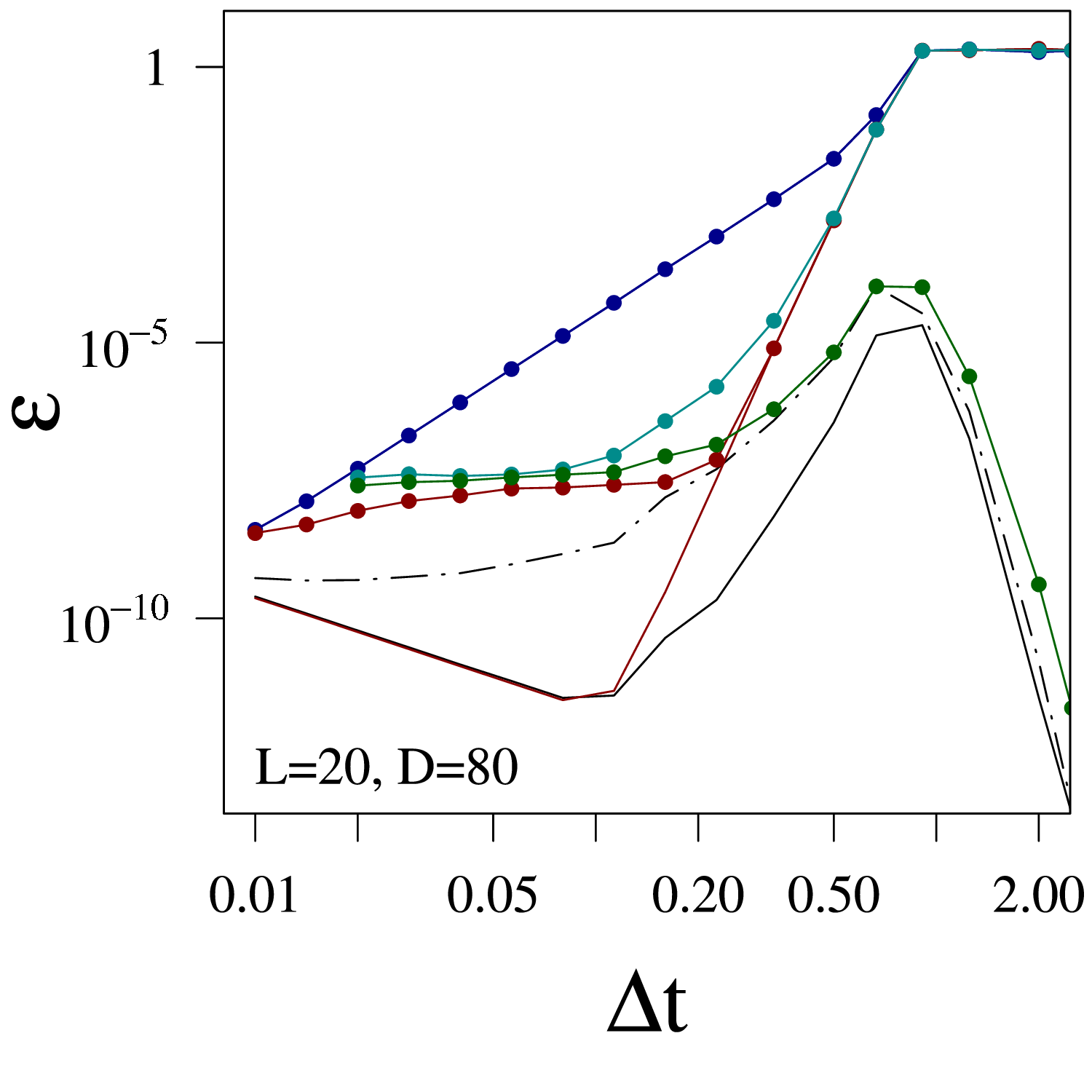}%
  \includegraphics[width=0.5\linewidth]{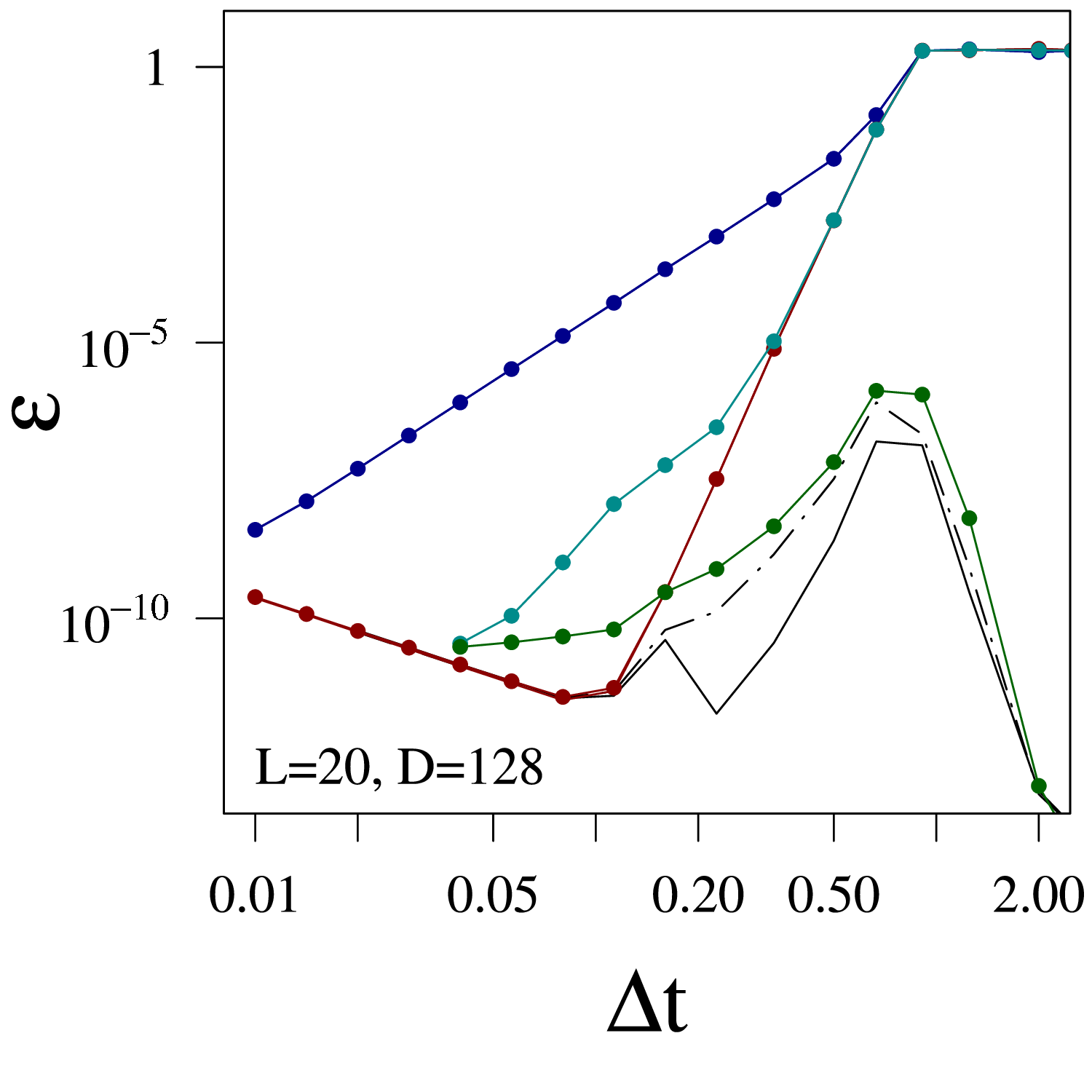}
\caption{Same as in Fig.~\ref{fig-exact} but for $N=20$ spins.}
  \label{fig-truncated-20}
\end{figure}

We tested all algorithms by simulating the evolution of the same state
under a family of spin-$\case{1}{2}$ Hamiltonians with
nearest-neighbor interactions
\begin{equation}
  \label{Hamiltonian}
  H =  \sum_k \left[ \cos(\theta) (s^x_k s^x_{k+1} + s^y_k
    s^y_{k+1} + \Delta s^z_k s^z_{k+1})+ \sin(\theta) s^z_k\right].
\end{equation}
As initial state we take the product $|\psi(0)\rangle \propto
(|0\rangle + |1\rangle)^{\otimes L},$ where $|0\rangle$ and
$|1\rangle$ are the eigenstates of $s^z.$ By restricting ourselves to
``small'' problems ($L\leq 20$), we can compare all algorithms with
accurate solutions based on exact diagonalizations and the Lanczos
algorithm \cite{noack05,hochbruck96}. Notice that we measure the error
in the full wavefunction, $\varepsilon := \Vert \psi_D(T) -
U(T)\psi(0)\Vert^2$ and not on the expectation values of simple
correlators whose exact evolution is known \cite{manmana06}. The
outcome of some of the simulations is in Figs.~\ref{fig-exact},
\ref{fig-truncated} and \ref{fig-truncated-20}, which we will discuss
in the following paragraphs.

The first set of simulations was done for 8 spins using matrices of
size $D=16$ and a DMRG basis of similar size. Since $S_{D}$ contains
all possible states, ${\cal P}_{D} = 1$, we expect no truncation
errors in any of the algorithms. As shown in Fig.~\ref{fig-exact} for
the XY model with $\theta=0.35$ and $\Delta=0,$ for medium to long
time steps most errors show the expected behavior. Thus, the errors of
the Trotter methods of second and fourth order follow the laws ${\cal
  O}(\Delta t^2)$ and ${\cal O}(\Delta t^4)$. Runge-Kutta, Taylor and
Pad\'e approximations have an error of ${\cal O}(\Delta t^2)$, and for
the Arnoldi and Lanczos methods with $N_v=4$ and 8 vectors we have a
qualitative behavior ${\cal O}(\Delta t^{N_v-1}).$ Since the size of
the matrices and of the DMRG basis is very big, all these laws are
followed, irrespective of whether the implementation uses DMRG or
MPS. The only exception seems to be the DMRG Lanczos when implemented
with only two target states. This method behaves more poorly than the
counterpart with $N_v$ target states given by
$\ket{\phi_k}:=\ket{\psi(k\Delta t/(N_v-1)},\,k=0..N_v-1$.

Out of the methods that work as expected, all of them break the ideal
laws at some point, acquiring an error of order ${\cal O}(\Delta
t^{-2}).$ This error is exponential in the number of steps $T/\Delta
t$ and it signals the finite accuracy of the optimization algorithms,
due to the limited precision of the computer. Roughly, since current
computers cannot compute the norms of vectors, $\Vert \psi\Vert^2,$
with a relative error better than $10^{-16},$ a worst case estimate is
$\epsilon = (T/\Delta t)^2 10^{-16}$, which perfectly fits these
lines.

Theoretically, the performance measured in computation time and memory
use of all algorithms is of order ${\cal O}(N_v N_H D^3/N),$ where
$N_H$ is related to the number of operators in $\{X, Y, X^\dagger X,
Y^\dagger Y,\ldots\},$ $N$ is the size of the problems and the
additional factor $N_v$ only appears for Arnoldi and Lanczos
methods. For periodic boundary conditions the cost increases by ${\cal
  O}(D^2).$ However, we have explicitely avoided PBC problems because
DMRG cannot handle them so accurately. In practice, out of all
methods, the Pad\'e expansions are the slowest ones, while the Trotter
formulas, being local, are the fastest. In between we find the Arnoldi
and Lanczos methods, which are nevertheless competitive if we consider
their accuracy and the fact that they allow for longer time steps.

In the remaining simulations we dropped the methods from
Sect.~\ref{sec:rkmps} and \ref{sec:pade}, because they have a similar
computational cost and worse performance than the MPS Arnoldi
method. We keep, on the other hand, all Trotter and DMRG methods and
continue our study with bigger problems in which the MPS spaces and
the DMRG basis are smaller than the limit required to represent all
states accurately, $D=d^{N/2}$ and $M=d^{N/2-1}$. More precisely, we
choose $N=16$ and $20$ spins and try with $M,D=32,64,80$ and $128.$

Now that the methods are potentially inexact, our previous error laws
are modified by the introduction of truncation errors. The first thing
we notice in Fig.~\ref{fig-truncated}a-c is that the Trotter error of
second order is rather stable, its error being the same for MPS and
DMRG. However, when we go for higher orders and increase the number of
exponentials, truncation errors affect more strongly the Vidal and
DMRG implementations than the MPS one. This shows the difference
between making local truncations after each bond unitary (DMRG and
Vidal) versus delaying truncations until the end and using the optimal
projection \cite{verstraete04b}.

Another important conclusion is that all DMRG methods are very
sensitive to truncation errors. As shown in
Figs.~\ref{fig-truncated}a-c, there is not a very big difference in
accuracy between the DMRG Runge-Kutta and the Lanczos implementation,
and both methods are not more accurate than the DMRG Forest-Ruth
formula. The main reason why higher order methods from
Sect.~\ref{sec:rk} do not improve the results is due to estimating the
evolution with the matrix of the Hamiltonian on the truncated DMRG
basis. Another reason is that under truncation, the Lanczos recurrence
does no longer produce an orthogonal set of vectors.

To prove those statements we compared with a Lanczos method
implemented with MPS as explained in Sect.~\ref{sec:arnoldi}. This
method does indeed have a better accuracy than the DMRG ones, which
can be attributed to the use of the full Hamiltonian. The errors of
the Lanczos are however larger than those of the Arnoldi method for a
similar $D$ and we have checked that under truncation the vectors of
the Lanczos basis are not truly orthogonal. These small errors
accumulate for shorter time steps, and only the Arnoldi method can
correct them.

\begin{figure}[t]
  \includegraphics[width=\linewidth]{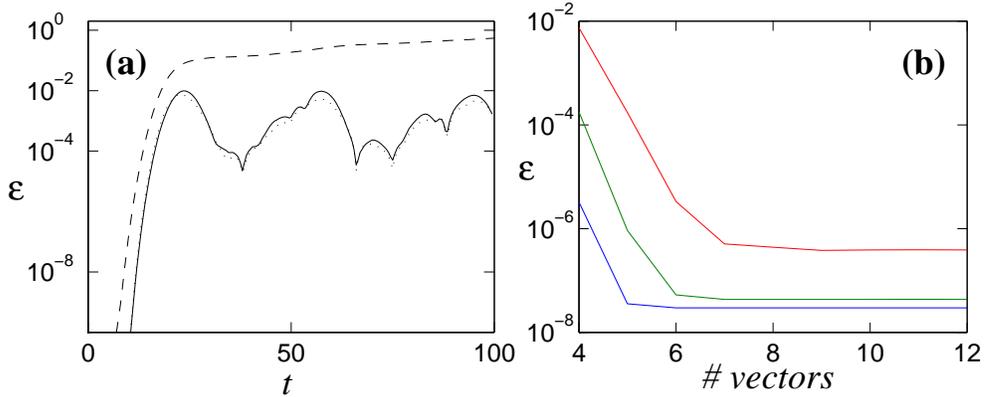}%
\caption{Simulations of $L=16$ spins with $\theta=0.35$,
  $\Delta=0$. (a) Minimal truncation error $\varepsilon=\Vert (1-{\cal
    P}_{64}\exp(-iHt)|\psi(0)\rangle\Vert^2$ of the time evolved state
  and accumulated error made when simulating this Schr\"odinger
  equation using the Arnoldi method with 8 vectors, $D=64$ and $\Delta
  t=0.16$ (dashed). (b) Similar as before, errors in the Arnoldi
  method for varying number of vectors, $T=10$ and $\Delta t= 0.04,
  0.16, 0.32$ (bottom to top).}
  \label{fig-truncation}
\end{figure}

As for the Arnoldi method, it has the greatest accuracy and seems to
be stable as $D$ becomes small. For very small matrices the error
remains constant as we decrease the time step, but this is only
because there is a lower bound in the approximation error given by
$\Vert{\cal P}_D\ket{\psi(T)} - \ket{\psi(T)}\Vert^2$.
Figure~\ref{fig-truncation}a shows how the errors in the Arnoldi
method are correlated to the errors made when approximating the exact
solution with a MPS of fixed size. This plot illustrates the fact that
all errors in this method are due to the final truncation.

Summing up, one should use the method that allows for the longest time
steps and the least number of truncations (or applications of ${\cal
  P}_D$) and mathematical operations. All methods have an optimal time
step which is a compromise between the errors in $U_n$ and the
rounding and truncation errors made on each step.  Regarding
performance and accuracy, the two winning methods are MPS algorithms
using either the fourth order Forest-Ruth decomposition or the Arnoldi
basis. The last method however, has two advantages. One is that it can
deal with nonlocal interactions and the second one is its potential
for parallelizability, roughly ${\cal O}(N_vN_H/L),$ which all other
presented algorithms lack. This can make it competitive with, for
instance, increasing the size of the matrices in the Forest-Ruth
method. Regarding DMRG methods, we find that they give comparable
results only for big basis. When truncation errors pop in, their
behavior is less predictable and it does not seem worth going with
more elaborate algorithms (Lanczos, Runge-Kutta) vs. an ordinary
Trotter formula.

\begin{figure}[t]
  \includegraphics[width=0.5\linewidth]{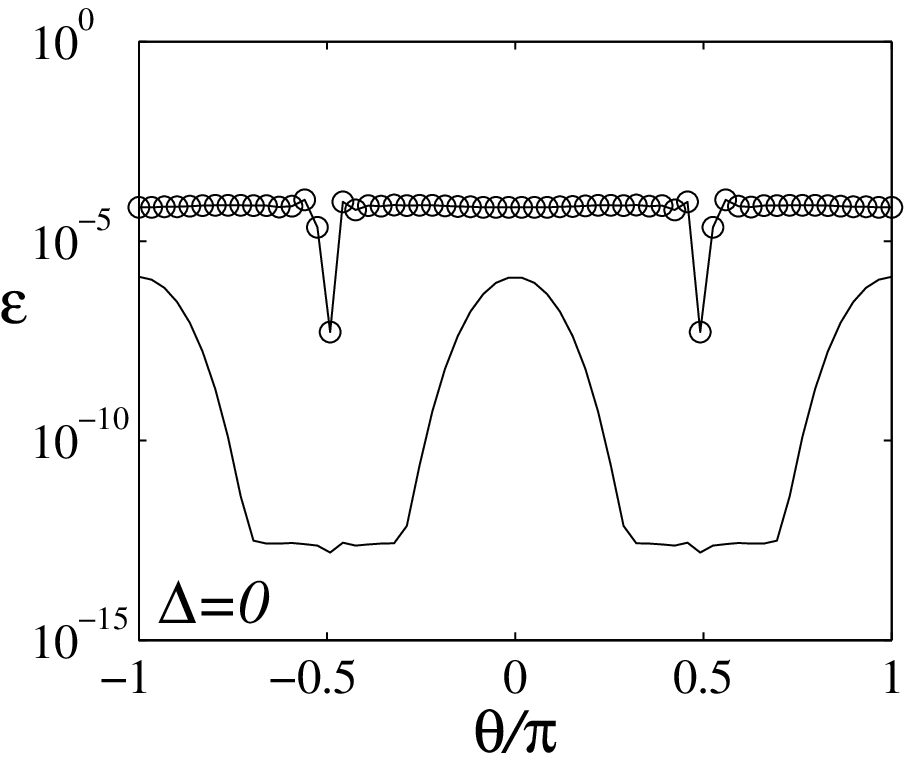}%
  \includegraphics[width=0.5\linewidth]{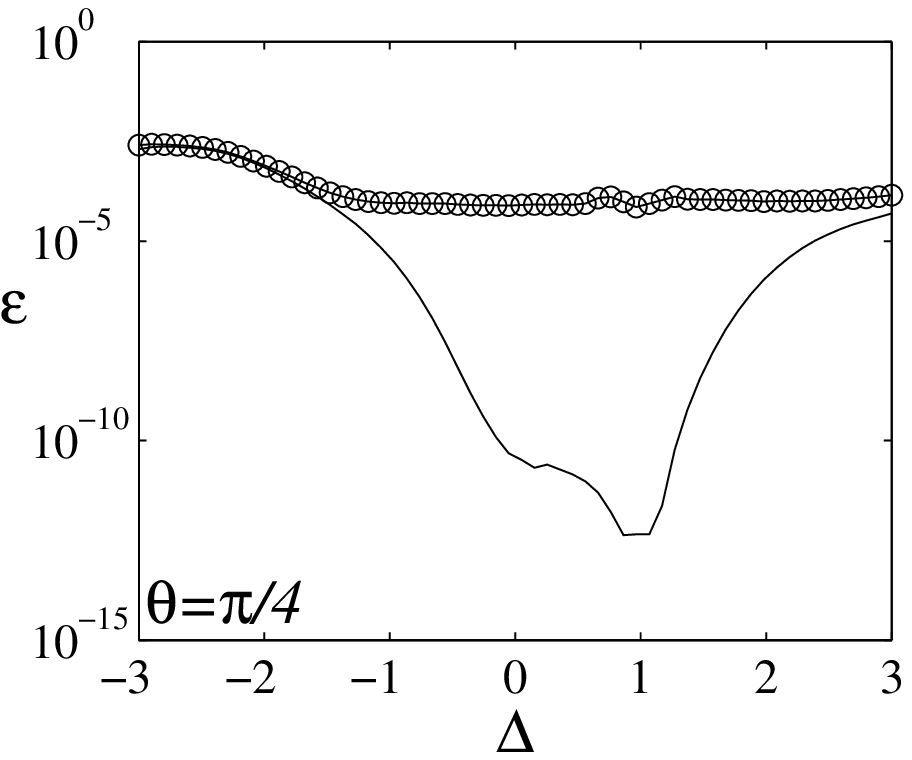}
\caption{Errors for various other spin $s=1/2$ models as parametrized
  in Eq.~(\ref{Hamiltonian}). We have used a second order Trotter
  method (circles) and an Arnoldi method (solid), with 16 spins, $T=5$
  and $D=32$.}
  \label{fig-scan}

\end{figure}

The fact that previous results are model-independent has been
confirmed by a systematic scanning of all possible Hamiltonians in
Eq.~(\ref{Hamiltonian}). A selection is shown in
Fig.~\ref{fig-scan}. The Arnoldi method is shown to be accurate,
even for gapless problems. When the Arnoldi method fails it is due to
truncation errors. In those situations the evolved state cannot be
accurately represented by MPS (and by that matter also not by DMRG),
simply because $\Vert(1-{\cal P}_D)\ket{\psi(t)}\Vert^2$ is finite and
large [Fig.~\ref{fig-exact}d]. Increasing the number of Arnoldi
vectors will also not help (See Fig.~\ref{fig-truncation}b) for the
same reason.

\section{Simulation of Feschbach resonances}
\label{sec:atoms}

As a real-world application we have used the MPS Trotter and Arnoldi
methods to simulate the conversion of bosonic atoms into molecules,
when confined in an optical lattices and moving through a Feschbach
resonance \cite{thalhammer06,stoferle06,volz06}. The goal is to study
how correlation properties are transferred from the atoms into the
molecules and how this dynamics is affected by atom motion and
conversion effic1iency.

The effective model combines the soft-core Bose-Hubbard model used to
describe the Tonks gas experiments \cite{paredes04}, with a coupling to
a molecular state \cite{dickerscheid05}
\begin{eqnarray}
\fl
  H = -J \sum_{\langle i,j\rangle,\,\sigma} a^\dagger_{i\sigma} a_{j\sigma}
  + \sum_{i,\sigma,\sigma'}\frac{U_{\sigma,\sigma'}}{2}
  a^{\dagger}_{i\sigma} a^{\dagger}_{i\sigma} a_{i\sigma} a_{i\sigma}
  \label{Hubbard}
  \\
  + \sum_i \left\{ (E_m + U_m n^{(a)}_i) n^{(m)}_i +
    \Omega [b^\dagger_i a_{i\uparrow} a_{i\downarrow} +
    H.~c.]\right\}.
  \nonumber
\end{eqnarray}
Here, $a_{i\uparrow},$ $a_{i\downarrow}$ and $b_i$ are bosonic
operators for atoms in two internal states and the molecule; $n^{(a)}$
and $n^{(m)}_i$ are the total number of atoms and of molecules on each
site, and we have the usual two-level coupling with Rabi frequency
$\Omega$ and detuning $\Delta := E_m - U_m.$  For simplicity, we will
assume that atoms and molecules interact strongly among themselves
($U_{\uparrow,\uparrow},U_{\downarrow,\downarrow},U_{m} \to \infty$),
so that we can treat them as hard-core,
$a_{i,\sigma}^2,b_i^2,a_{i\sigma}b_i=0.$ Also since molecules are
heavier, we have neglected their tunneling amplitude, although that
could be easily included.

\begin{figure}[t]
  \includegraphics[width=\linewidth]{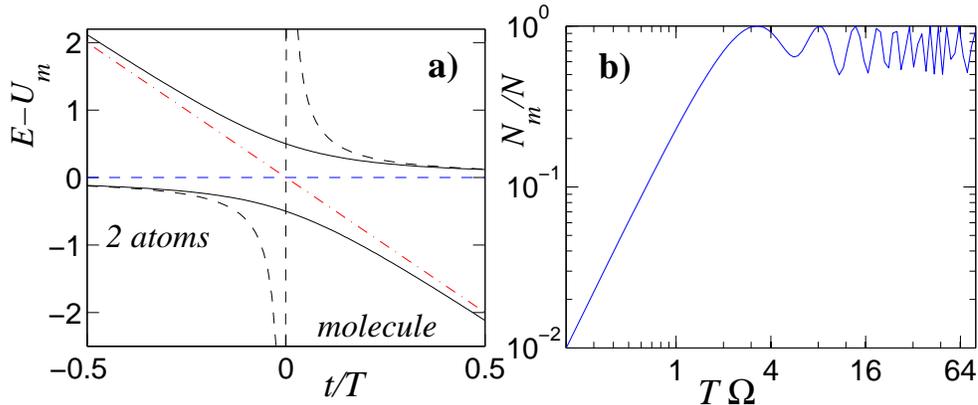}%
  \caption{Dynamics of a site with two atoms when the energy of the
    molecules is ramped linearly: $E_m = U_{\uparrow\downarrow} +4
    \Omega (1 - 2t/T).$ We plot (a) the instantaneous energy levels
    (solid), and (b) the fraction of atoms converted into molecules.}
  \label{fig-molec-L2}
\end{figure}

As the energy of the molecular state is shifted from $E_m \gg U_m$
down to $E_m \ll U_m,$ the ground state of Eq.~(\ref{Hubbard}) changes
from a pair of coupled of Tonks gases, to a purely molecular
insulator. We want to study the dynamics of this crossover as $E_m$ is
ramped slowly from one phase to the other.

The simplest situation corresponds to no hopping: isolated atoms
experience no dynamics, while sites with two atoms may produce a
molecule. The molecular and atomic correlations at the end of the
process are directly related to two-body correlations in the initial
state \cite{altman05,barankov05},
\begin{eqnarray}
  \langle m_k^\dagger m_k\rangle_{t=T} &\sim& \langle n_{k\uparrow}
  n_{k\downarrow}\rangle_{t=0},\label{corr}\\
  \langle a_k^\dagger a_k\rangle_{t=T} &\sim&
  \langle a_k^\dagger a_k\rangle_{t=0} - \langle n_{k\uparrow}
  n_{k\downarrow}\rangle_{t=0}.
\end{eqnarray}
Therefore, this process can thus be used as a tool to probe quantum
correlations between atoms. Studying the two-level system
$\{a^\dagger_{k\uparrow}a^{\dagger}_{k\downarrow}|0\rangle,
b_k^\dagger|0\rangle\},$ we conclude that for this process to work
with a $90\%$ efficiency, the ramping time should be larger than
$T\sim 1.5/\Omega$ [See Fig.~\ref{fig-molec-L2}(b)].

\begin{figure}[t]
  \includegraphics[width=\linewidth]{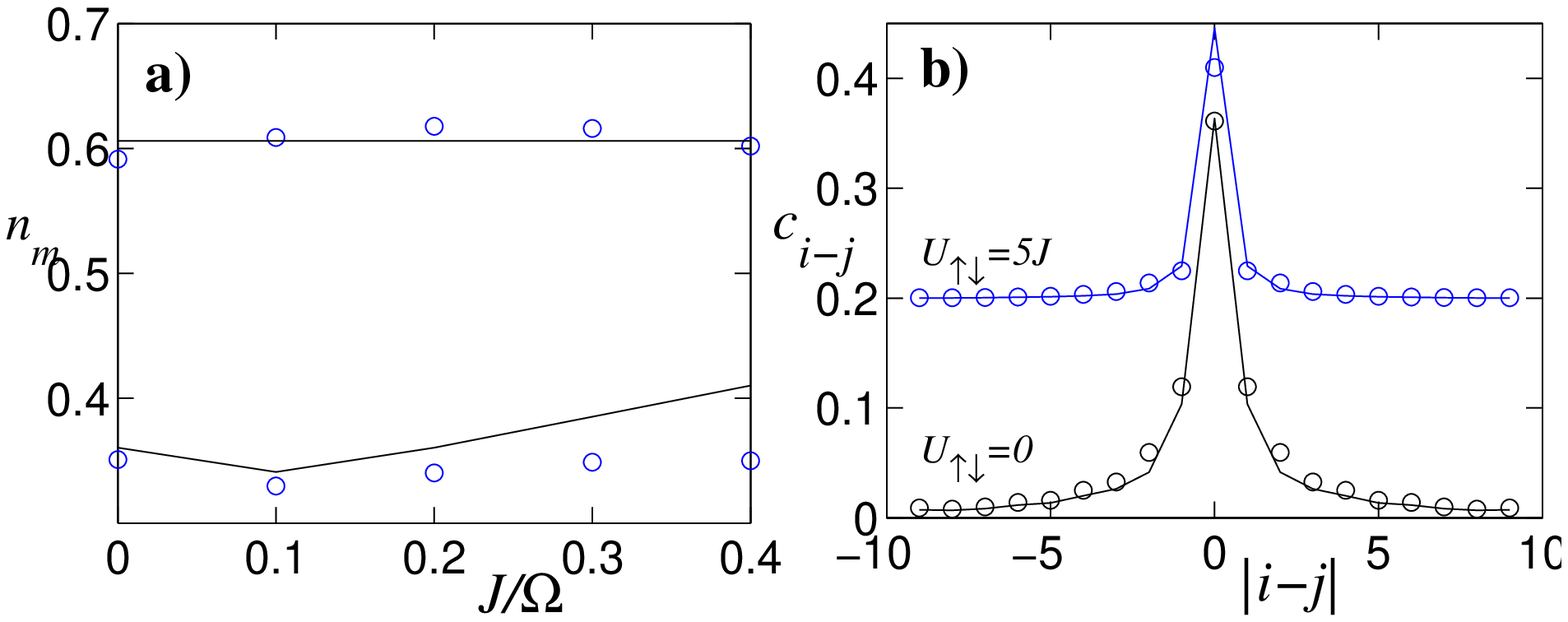}%
  \caption{(a) Fraction of atoms converted into molecules vs.
    adimensionalized hopping amplitude, for
    $U_{\uparrow,\downarrow}=0$ and $U_{\uparrow\downarrow}=1,$ from
    top to bottom. Circles show the outcome of the numerical
    experiment, while solid lines contain the ideal fraction
    (\ref{corr}). The case $J=0$ uses an initial condition
    $J=0.1\Omega$ and then switches off tunneling before ramping. (b)
    Correlations of the molecular state, $\langle m_i^\dagger
    m_j\rangle_{t=T}$ (circles) after the ramp, and those of the
    initial atomic state, $\langle
    a^\dagger_{i\downarrow}a^\dagger_{i\uparrow}
    a_{j\uparrow}a_{j\downarrow}\rangle_{t=0}$
    (solid line).  The plot for $U_{\uparrow\downarrow}=5J$ has been
    shifted up by $0.2.$}
\end{figure}

We have simulated numerically the ramping of small lattices, $L=10$ to
$32$ sites, with an initial number of atoms $N_{\uparrow,\downarrow} =
L/2, 3L/4.$ The value of the molecular coupling has been fixed to
$\Omega = 1$ and the interaction has been ramped according to $E_m =
U_{\uparrow\downarrow} + 4 \Omega (1 - 2 t/T)$ using the ideal ramp
time $T=1.5/\Omega.$ We have used two particular values of the
inter-species interaction, $U_{\uparrow,\downarrow}/\Omega = 0,2,$ and
scanned different values of the hopping $J/\Omega\in [0,0.4].$ The
initial condition was always the ground state of the model with these
values of $U_{\uparrow\downarrow}/J$ and no coupling. These states
contain the correlations that we want to measure.

The main conclusions is that indeed the correlations of the molecules
are almost those of the initial state of the atoms (\ref{corr}), even
for $J=0.4\Omega$ when the process has not been adiabatic. An
intuitive explanation is that hopping is strongly suppressed as we
approach the resonance, due to the mixing between atomic states, which
can hop, and molecular states, which are slower. We can say that the
molecules thus pin the atoms and \textit{measure} them. This
explanation is supported by a perturbation analysis at $J\ll \Omega,$
where one finds that a small molecular contamination slows the atoms
on the lattice. This analysis breaks down, however, for $J \sim
\Omega,$ the regime in which the numerical simulations are required.

\section{Conclusions}
\label{sec:conclusions}

We have performed a rather exhaustive comparison of different methods
for simulating the evolution of big, one-dimensional quantum systems
\cite{vidal04,verstraete04b,white93,daley04,gobert05,feiguin04,schmitteckert04,manmana05,manmana06}
with three other methods developed in this work. We find the MPS
methods to be optimal both in accuracy and performance within the
formulas of similar order. All procedures are substantially affected
by truncation and rounding errors, and to fight the latter we must
choose large integration time-steps. However, the only algorithm which
succeeds for very large time-steps is an Arnoldi method developed in
this work. Finally, this algorithm can be applied to problems with
long range interactions.

Using this algorithm, we have simulated the dynamics of cold atoms in
a 1D optical lattice when crossing a Feschbach resonance. The main
conclusion is that with rather fast ramp times it is possible to map
the correlations of the atomic cloud (two Tonks gases in this case)
and use this as a measuring tool in current experiments. This result
connects with similar theoretical predictions for fermions in
Ref.~\cite{altman05,barankov05}. Simple generalizations of this work
will allow us in the future to analyze losses and creation of strongly
correlated states with the help of the molecular component.

As posible outlook, we envision the possibility of developing new
algorithms in which the state is approximated by a linear combination
of MPS at all times. This should be more efficient than increasing the
size of the matrices, and could support distributed computations in a
cluster.

\section*{References}

%\bibliographystyle{unsrt}
%\bibliography{atoms}

\end{document}